\documentclass{aa}
\usepackage{graphicx}
\usepackage{txfonts}
\usepackage{natbib}
\usepackage{color}

\bibpunct{(}{)}{;}{a}{}{,}

\begin{document}

\title{Hydrodynamical simulations of the jet\\ in the symbiotic star MWC 560}
\subtitle{I. Structure, emission and synthetic absorption line profiles}

\titlerunning{Hydrodynamical simulations of the jet in MWC 560. I.}

\author{Matthias Stute\inst{1} \and Max Camenzind\inst{1} \and Hans Martin 
Schmid\inst{2}}
\institute{Landessternwarte Heidelberg, K\"onigsstuhl, D-69117 Heidelberg, 
Germany \and Institute of Astronomy, ETH Zentrum, CH-8092 Zurich, Switzerland}

\offprints{Matthias Stute, \email{M.Stute@lsw.uni-heidelberg.de}}

\date{Received 25 May 2004 / Accepted 25 August 2004}

\abstract{
We performed hydrodynamical simulations with and without radiative cooling of 
jet models with parameters representative for the symbiotic system MWC 560. 
For symbiotic systems we have to perform jet simulations of a pulsed 
underdense jet in a high density ambient medium. We present the jet structure 
resulting from our simulations and calculate emission plots which account for 
expected radiative processes. In addition, our calculations provide expansion 
velocities for the jet bow shock, the density and temperature structure in the 
jet, and the propagation and evolution of the jet pulses.

In MWC 560 the jet axis is parallel to the line 
of sight so that the outflowing jet gas can be seen as blue shifted, variable 
absorption lines in the continuum of the underlying jet source. Based on our 
simulations we calculate and discuss synthetic absorption profiles. 

Based on a detailed comparison between model spectra and 
observations we discuss our hydrodynamical calculations for a pulsed jet in 
MWC 560 and suggest improvements for future models.
}

\maketitle

\section{Introduction}

Jets are very common in a variety of astrophysical objects on very different 
size and mass scales. They can be produced by supermassive black holes in the 
case of Active Galactic Nuclei (AGN), by stellar black holes in Black Hole 
X-ray Binaries (BHXBs), by neutron stars in X-ray Binaries, by pre-main 
sequence stars in Young Stellar Objects (YSO) and by white dwarfs in supersoft 
X-ray sources and in symbiotic binaries. Jets in symbiotic systems are not yet 
as well investigated theoretically as the other objects. Due to the fact, that
their parameters are in a different regime, studying them should promise 
new insights. The density contrast $\eta = \rho_{\rm{jet}}/\rho_{\rm{ambient}}$
is about $10^{-3}-10^{-2}$ in symbiotic stars ($1-10$ in YSO, $\approx 0.1$ in 
AGN), the outflow velocities are with 1000-5000 km s$^{-1}$ 
somewhat higher than in YSOs with $100-1000$ km s$^{-1}$ ($\approx c$ in AGN).
The absolute densities of ${\rm N}_{e} \approx 10^6-10^8$ 
cm$^{-3}$ are similar to YSO jets. However, the jet gas 
densities in symbiotic systems are the highest for 
{\em underdense} jets ($\eta < 1$) -- which are in AGN jets smaller than 
$10^{-2}$ cm$^{-3}$ -- therefore radiative processes become very important. 

Symbiotic systems consist of a red giant undergoing strong mass loss and
a white dwarf. More than one hundred symbiotic stars are 
known, but only about ten 
systems show jet emission. The most famous systems are R Aquarii, CH Cygni and 
MWC 560. While the first two objects are seen at high inclinations -- a fact 
which makes it possible to study the morphology and structure of jets of 
symbiotic stars -- the jet axis in MWC 560 is practically parallel to the line 
of sight. This special orientation provides the opportunity to observe the 
outflowing gas as line absorption in the source spectrum. With such 
observations the radial velocity and the column density of the outflowing jet 
gas close to the source can be investigated in great detail. In particular we 
can probe the acceleration and evolution of individual outflow components with 
spectroscopic monitoring programs as described in \citet{SKC}. 

MWC 560 is a symbiotic binary system with a late M giant undergoing strong 
mass loss. At least a significant fraction of the lost material is accreted
by the companion, which is as for most symbiotic stars a white dwarf. No 
radial velocity variation have been detected for the red giant most likely 
because the inclination of the system is close to $0^{\circ}$. The orbital 
period of the system is not known. However, \citet{SKC} provide arguments for 
a likely orbital period in the range 4 to 10 years.

We performed hydrodynamical simulations with and without cooling of jets with 
parameters that are intended to represent those in MWC 560. In a grid of eight
simulations we investigated the influence of different jet pulse parameters. 
Due to the high computational costs of simulations including cooling, this 
grid was restricted to adiabatic simulations. Existing simulations of pulsed 
jets for YSO systems 
\citep[e.g.][and references therein]{StN,SGW,GPC} showed that the resulting 
jet structure differs strongly between purely hydrodynamical models and models 
using radiation hydrodynamics. Therefore, one model simulation was 
performed which includes a treatment of radiative cooling. We calculate from 
these models the absorption line profiles and investigate their ability to 
explain the corresponding spectroscopic observations. 

Due to the fact that morphological studies of the jet structure and synthetic 
emission plots are the first model results which are obtained, we also present 
them in this paper. In addition, our calculations provide expansion velocities 
for the jet bow shock, the density and temperature structure in the jet, and 
the propagation and evolution of the jet pulses. This allows us to compare at 
least the qualitative properties of the jet simulations with various types of 
observations of jets in symbiotic systems. 

In section \ref{sec_obs}, we summarize the main observational results of the 
jet sources in symbiotic systems and in particular for the jet absorptions
in MWC 560. A detailed description of our hydrodynamic model is given in 
section \ref{sec_models}. The resulting jet structures are described and 
emission plots are presented (section \ref{sec_struc}). In section 
\ref{sec_line_param}, we discuss the model parameters which define the 
synthetic absorption line profiles and the results for different model cases. 
Section \ref{sec_time_var} describes the temporal evolution of the high 
velocity gas in the pulses and compares the resulting jet absorption profiles 
with observation. Finally a summary and a discussion are given.

\section{MWC 560 and jets in symbiotic stars} \label{sec_obs}

Symbiotic binaries are a very heterogeneous class of objects, showing 
different types of nova-like activity. Jets are only expected in systems with 
substantial accretion from the red giant via a disk. The presence of a disk 
can be inferred from strong, short term ($\sim$ hour) flickering of the hot 
component. However, flickering is only observed in very few objects 
\citep{SBH}. In many systems an accretion disk may be present,
but it is hard to observe due to the 
much stronger emission from the cool giant, the nebula or the accreting white 
dwarf. Further the disk may be hidden by a larger scale circumbinary disk. In 
some systems the jet emission seems to be a transient phenomenon
\citep[e.g.][]{TMM}, connected 
with an active phase of the hot component - perhaps due to an accretion disk 
instability. However, such transient features are difficult to study 
observationally.

\subsection{MWC 560}

The jet in the symbiotic system MWC 560 serves in this study as template 
object for our model calculations. We have chosen this unique jet system 
because it provides us with direct information on some hydrodynamical 
parameters of the jet gas in the near vicinity $<1$~AU of the jet source. 

MWC 560 is observationally a point source. In early 1990, MWC 560 attracted 
attention with a photometric outburst of 2 mag. With spectroscopic 
observations it was found that the system exhibits strongly variable, blue 
shifted absorptions with outflow velocities (RV) up to 6000 km s$^{-1}$ 
\citep{TKG}. Unlike in normal P Cygni profiles from a stellar wind the blue 
absorption components are detached from the emission component. After the 
initial outburst phase the outflow showed during about a year a low velocity 
phase with $v=300\,{\rm km}\,{\rm s}^{-1}$. Since September 1991, the 
``normal'' outflow mode is re-established, with strongly variable, detached 
absorption components and a typical outflow velocity of $\approx$ 1500 km 
s$^{-1}$ \citep{ToK}. The absorption line structure can be explained as a jet 
outflow whose axis is parallel to the line of sight \citep{TKG, SKC}. 
This very special system orientation is supported by the absence of 
measurable radial velocity variations for the red giant indicating that the 
orbital plane and therefore presumably also the accretion disk are 
perpendi\-cular to the line of sight. Moreover strong flickering is present 
\citep{SBH} as 
expected for an accretion disk of a strong jet source seen pole on. Up to now, 
this object is the only stellar object known with this special jet 
orientation. Therefore MWC 560 is most useful for studying the acceleration and
dynamical evolution of small scale structures in a stellar jet. Studying the 
variable gas absorptions yields information about the outflowing gas at very 
small distances $< 10$ AU from the source. 

\begin{figure}
   \resizebox{\hsize}{!}{\includegraphics{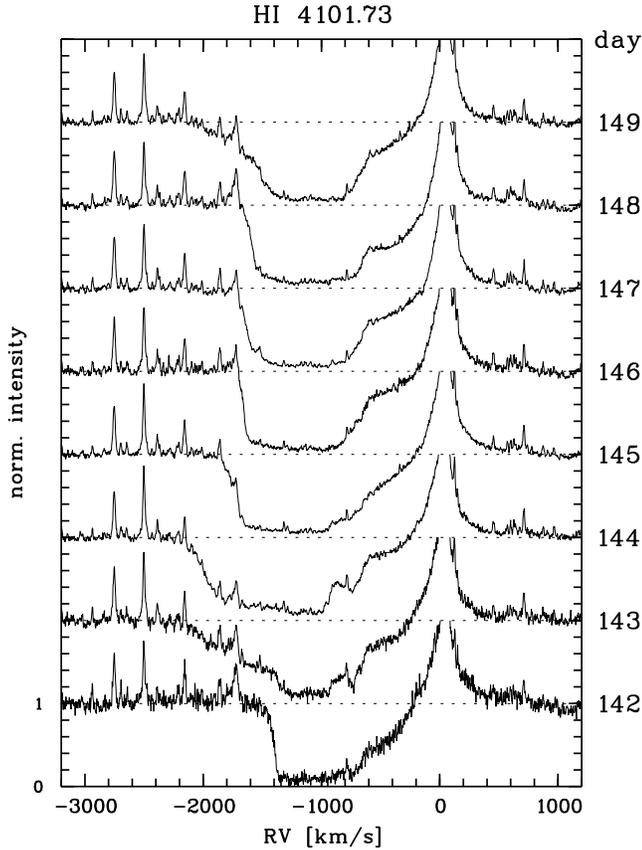}}
   \caption{Time series for the H$\delta$ $\lambda$4102 jet absorption
     in MWC 560 from the campaign of \citet{SKC}. The
     sequence shows normalized spectra (and shifted for clarity) from
     days before and after the appearance of a strong high 
     velocity component on day 144.}
   \label{hmsprof}
\end{figure}

A most important observational source of information for the investigation of 
the pulsed jet in MWC 560 are the monitoring observations described in 
\citet{SKC}. Figure \ref{hmsprof} displays a small fraction of the H$\delta$ 
jet absorption data obtained during this campaign. It is clearly visible that 
the jet absorptions in MWC 560 are very different from classical P Cygni 
profiles of stellar winds. The variable absorption features can be grouped 
into three different components (see Fig. \ref{hmsprof}).

\begin{itemize}
\item{}The first component is relatively stable with RV of 
  \mbox{$\approx -1200 \pm 300$ km s$^{-1}$}. This component is saturated over 
  a wide velocity range in the \ion{H}{i} Balmer lines, while the metal lines 
  (\ion{Na}{i}, \ion{Fe}{ii}, \ion{Ca}{ii}) are not (or not completely)
  saturated and several subcomponents can be resolved. Focusing specifically 
  on the \ion{Na}{i} line, one notices that no absorptions with RV 
  \mbox{$\ge -1000$ km s$^{-1}$} are present. This shows that low velocity 
  gas is at least partly ionized and produces no 
  \ion{Na}{i} absorptions. 
\item{} The second component are transient high-velocity absorptions
  which appear repeatedly on timescales of about a week. In Fig. \ref{hmsprof} 
  such a strong high velocity absorption appears on day 143 and is fully 
  developed on day 144. For day 144 the maximum outflow 
  velocity is about \mbox{$2100$ km s$^{-1}$} while other high velocity 
  components reach velocities up to  \mbox{$2600$ km s$^{-1}$} 
  \citep[days 181 and 193, see][]{SKC}. These fast components can be seen in 
  all \ion{H}{i} lines and simultaneously in the metal lines. Particularly 
  strong, compared to the stable component around 
  \mbox{$\approx -1200 \pm 300$ km s$^{-1}$}, are the high-velocity components 
  in \ion{He}{i}. The transient high velocity components reach their maximum 
  velocity within one day and decay again in the following few days. The 
  disappearance is faster in \ion{He}{i} than in \ion{H}{i}. The \ion{He}{i} 
  transitions have a higher excitation potential, therefore the \ion{He}{i} 
  features can be associated with hotter and/or higher ionized gas. 
\item{} The third absorption component in the RV range of 
  \mbox{$\approx -900$ to $-400$ km s$^{-1}$} weakens at the same time
  as the appearance of the high velocity component. In Fig. \ref{hmsprof}
  the reduced absorption (enhanced emission) is particularly well visible for 
  days 143 and 144 around \mbox{$-900$ km s$^{-1}$}. The low velocity 
  absorptions are anti-correlated with the transient high velocity components. 
  This is an indication for a close relationship between the acceleration 
  region producing the low velocity absorption and the transient high velocity 
  components.
\end{itemize}
The observations provide much more information on the jet structure in MWC 560 
and we refer the interested reader to \citet{SKC}. 

In that work, various jet parameters for MWC 560 were 
estimated from the observations.Of interest for this study 
on synthetic spectra from hydrodynamic models is the estimated jet mass 
outflow rate of $>7\times 10^{-9}\,{\rm M}_\odot\,{\rm yr}^{-1}$. 
In addition it was possible to derive values for the velocity 
and the gas density for the ``normal'' jet outflow, but also for phases with 
strong high velocity components. These determinations of hydrodynamical 
parameters of the initial jet gas in MWC 560 serve now in this paper as input 
parameter for the numerical simulations of pulsed jets in symbiotic systems.

The pole on orientation of the jet in MWC 560 has of course the drawback that 
this system is not suited for investigations on the morphological structure of 
the jet. For this we have to use observations of jets which are seen from the 
side. 

\subsection{R Aquarii and CH Cygni}

R Aquarii is with a distance of about 200 pc one of the nearest symbiotic 
stars and a well known jet source. The system contains a Mira-like variable 
with a pulsation period of 387 days. The hot, ionizing companion is not 
resolved, but it is presumably a white dwarf or sub-dwarf with an accretion 
disk. An orbital period of $\approx 44$ years has been suggested, but this 
value is highly uncertain \citep{Wal}. The jet has been extensively observed 
in the optical, at radio wavelengths and with X-ray observations 
\citep[e.g.][]{SoU,PaH,HMKM,HLDF,KPL}. R Aqr shows a jet and a 
counter-jet extending about 10$''$ each. The jets are embedded in an extended 
and complex nebulosity. Individual jet features are morphologically and 
photometrically variable with time. In observations with HST \citep{PaH}, the 
jet can be traced down to a distance of only 15 AU from the Mira where it is 
already collimated with an opening angle of $< 15{^\circ}$. After a straight 
propagation of 50 AU, it hits a dense clump and produces a radiative bow shock 
\citep[feature N2; notation as in][]{PaH}. The flow seems to split 
into two parts: one stream extends around 700 AU towards feature A1, further a 
series of parallel features (N3 - N6) are detected downstream of N2, 
orthogonal to the original flow. HST observations taken at different epoch 
revealed that the transverse velocity (proper motion) of the different 
features increases from about 40 to 240 km s$^{-1}$ with increasing distance 
from the central jet \citep{HLDF}. In Chandra images \citep{KPL}, the knots 
are not as well resolved as in optical observations. Only larger clumps are 
visible, corresponding to the central source, the feature A1 and the feature 
S3 in \citet{PaH}. VLA observations \citep{HMKM} show similar structures. 

For R Aqr a major problem for the interpretation of the jet observations is 
the blending of jet features with emission from the surrounding nebulosity. 
Therefore it is not clear whether the observed features are due to the jet gas,
due to the ambient medium which is shock-excited by the jet outflow, or just 
circumstellar material ionized by the radiation from the accreting component. 

CH Cygni is a symbiotic binary where the hot component shows strong, short term
flickering as expected for a bright accretion disk. The cool companion is an 
extended M6 III giant with a radius of about 200 $\mbox{R}_{\odot}$. The 
binary period is 5700 d \citep{CDS} and eclipses of the flickering component 
indicate a system inclination near 90$^{\circ}$ \citep{MMT}. In 1984/85, the 
system showed a strong radio outburst, during which a double-sided jet with 
multiple components was ejected \citep{TSM}. This event enabled an accurate 
measurement of the jet expansion with an apparent proper motion of 1.1 arcsec 
per year. With a distance of 268 pc (HIPPARCOS) \citep{CDE}, this leads to a 
jet velocity near 1500 km s$^{-1}$. The spectral energy distribution derived 
from the radio observations suggest a gas temperature of about 7000~K for the 
propagating jet gas \citep{TSM}. In HST observations \citep{EBS}, arcs can be 
detected which also could be produced by episodic ejection events. The X-Ray 
spectrum taken with ASCA \citep{EIM} is composed of three components which are 
associated with the hot source and the secondary or the radio jet. X-Ray 
imaging is not yet available.

\section{The numerical models} \label{sec_models}

In the following, we describe the employed computer code with the incorporated 
equations, the model geometry and the chosen jet parameters. We also discuss 
the simplifications and approximations made due to the constraints set by the
available computer resources. 

\subsection{The computer code}

\subsubsection{General description}

With the code {\em NIRVANA} \citep{ZiY} we solve the following set of the 
hyperbolic differential equations of ideal hydrodynamics 
\begin{eqnarray} \label{hydro}
\frac{\partial\,\rho}{\partial\,t} + \nabla\,(\rho\,{\bf v}) &=& 0 \nonumber \\
\frac{\partial\,(\rho\,{\bf v})}{\partial\,t} + \nabla\,
(\rho\,{\bf v}\otimes{\bf v}) &=& - \nabla\,p - \rho\,\nabla\,\Phi \\
\frac{\partial\,e}{\partial\,t} + \nabla\,(e\,{\bf v}) &=& - p\,\nabla\,{\bf v}
 + \Lambda ( T ; \rho_{i}) \nonumber \\
p &=& (\gamma - 1)\,e .\nonumber 
\end{eqnarray}
\noindent
Thereby $\rho$ is the gas density, $e$ the energy density, ${\bf v}$ the 
velocity and $\gamma$ the ratio of the specific heats at constant pressure and 
volume. {\em NIRVANA} uses second order accurate finite-difference and 
finite-volume methods and explicit time-stepping. 

This code was modified by M. Thiele \citep{Thi} to calculate energy losses 
due to non-equilibrium cooling by radiative emission processes. The 
microphysics is introduced via the cooling term $\Lambda$ in the energy 
equation and the following interaction equations for the species. When cooling 
is important, the above equations are supplemented by a species network
\begin{equation} \label{interact}
\frac{\partial\,\rho_{i}}{\partial\,t} + \nabla\,(\rho_{i}\,{\bf v}) =
\sum_{i=1}^{N_{s}}\,\sum_{j=1}^{N_{s}}\,k_{ij} (T)\,\rho_{i}\,\rho_{j}
\end{equation}
\noindent
with $\rho_{i}$ the species densities satisfying $\rho = \sum_{i=1}^{N_{s}}\,
\rho_{i}$ for the total density. $k_{ij}$ are the rate coefficients for 
two-body reactions which are functions of the fluid temperature $T$. They 
describe electron collision ionization and radiative and dielectronic 
recombination processes. The summations go over the 
$N_{s}$ species. Both atomic and/or molecular species can be included in the 
model and {\em NIRVANA\_C} can handle up to 36 species \citep{Thi}.

In the optically thin case, the cooling rate is a function of the species 
densities and temperature, $\Lambda = \Lambda ( T ; \rho_{i})$ \citep{SuD}. 
Cooling functions describing electron collision ionization, radiative and 
dielectronic recombination and line radiation. When cooling is very efficient, 
the atomic network is solved in a time-implicit way \citep{Thi}. This code 
has been extensively tested in \citet{Thi} and \citet{Kra}.

When the cooling is solved dynamically with the full set of non-equilibrium 
equations, the various ionization states and concentration densities 
$\rho_{i}$ of each element are calculated from the atomic rate equations. They 
are used explicitly in the cooling functions as 
\begin{equation}
\Lambda ( T ; \rho_{i}) = \sum_{i=1}^{N_{s}}\,\sum_{j=1}^{N_{s}}\,e_{ij} (T)\,
\rho_{i}\,\rho_{j} + \Lambda_{BS} 
(T)
\end{equation}
\noindent
with $e_{ij}$ the cooling rates from two-body reactions between species $i$ and
$j$, and $\Lambda_{BS}$ the cooling function due to Bremsstrahlung. In this 
case, the equation of state has to be given in the form
\begin{equation}
T = \frac{\gamma - 1}{k_{B}}\,\frac{e}{\sum_{i=1}^{N_{s}}\,n_{i}} .
\end{equation}
\noindent
The code in its present form can handle collisional excitation, collisional 
ionization, recombination, metal-line cooling and Bremsstrahlung.

\subsubsection{Our cooling setup}

The available computer resources set strong constraints to the microphysics
which can be included in our models. Important for jet simulations is
a high spatial resolution in order to resolve the fine structure. 

The parameter study for pulsed jets is therefore restricted to adiabatic 
simulations, where we neglect both the cooling term $\Lambda$ in the energy 
equation in the set of equations (\ref{hydro}) and the interaction equations 
(\ref{interact}). 

In addition, one model simulation was performed which includes a simple 
treatment of radiative cooling. Instead of the full explicit cooling function 
$\Lambda ( T ; \rho_{i})$, we consider only cooling by hydrogen, together with 
a general non-equilibrium cooling function $\Lambda ( T )$ adapted from 
\citet{SuD} -- without radiation field and assuming solar abundances -- to 
account for the cooling by the heavier elements. The cooling function 
neglects collisional de-excitation. For the temperature 
regime of our calculations 
T $> 10000$ K collisional deexcitation is only of importance for densities 
above 10$^8$ cm$^{-3}$. Thus in the jet regions with the highest densities our 
cooling rate may be slightly overestimated. This effect will most likely be 
more than overcompensated by the fact that the limited resolution of our 
calculations underestimates the gas density (clumping) and the cooling rate. 
For hydrogen we solve the atomic network of \ion{H}{i}, \ion{H}{ii} and 
e$^{-}$ and calculate the cooling due to collisional ionization of \ion{H}{i} 
and due to collisional excitation of hydrogen line emission. The abundances of 
the species follow from the assumed initial jet temperature of $10^4$ K and 
the total density (see below).

\subsection{Geometric model}

\begin{figure}
   \resizebox{\hsize}{!}{\includegraphics{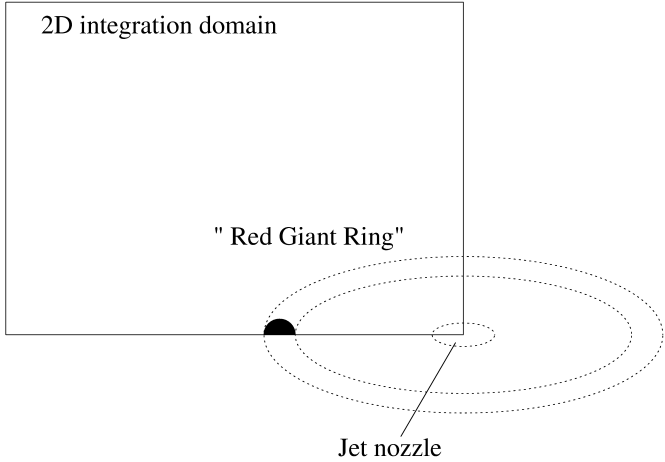}}
   \caption{Sketch of the reduction from 3D to 2.5D simulations}
   \label{skizze}
\end{figure}

Due to high computational costs of our cooling treatment and in order to 
combine a large computational domain with high spatial resolution 
it was not possible to perform model simulations in three dimensions. 
Therefore we had to choose a two-dimensional slice of the full domain and to 
assume axi\-symmetry. This kind of simulation is often called 2.5D 
simulations. The geometry of this slice in the examined system is shown in 
Figure \ref{skizze}. The dimensions of this two-dimensional slice are set to 
50 AU in polar direction perpendicular to the orbital plane of the binary and 
30 AU in the direction of the orbital plane -- Figure \ref{skizze} is not 
drawn to scale.

The hot component is located in the origin of the coordinate frame, therefore 
the companion is expanded into a ``Red Giant Ring''. In the 2D integration 
domain, only half of the cross-section of this ring on this slice is included. 
The binary separation in the models is chosen to 4 AU which is of the order of 
the estimated separations of 3.3 - 5.2 AU. The density of the red giant is set 
to $2.8 \cdot 10^{-5}$ g cm$^{-3}$ and its radius to 1 AU.

Surrounding the red giant, a stellar wind is implemented. The wind has a 
constant velocity of $v=10$~km/s, a gas temperature of $T=50$~K and a mass 
loss of 10$^{-6}$ $\mbox{M}_{\odot}$ yr$^{-1}$. The density of the red giant 
wind at the surface of the star is then $2.2 \cdot 10^{-14}$ g cm$^{-3}$. The 
density of the external medium is given by an $1/r^2_{\rm{rg}}$-law for a 
spherical wind where $r_{\rm{rg}}$ is the distance from the center of the 
red giant. The density of the red giant wind near the jet nozzle is about 200 
times higher than the initial jet density. At a distance of 50 AU from the 
symbiotic system, i.e. at the end of the integration domain, the wind density 
is about equal to the jet density at the nozzle.

In 3D this density distribution corresponds for small z ($<5$ AU) to a 
torus-like structure which becomes for large z ($>10$ AU) a slightly flattened 
but quasi-spherical distribution $\approx 1/r^2$ centered on the jet nozzle. 
This seems to be a reasonable approximation for the z-direction. In the 
r-direction we introduce by this procedure artifically a density symmetry 
around the jet axis, which may be a qualitatively important difference compared
to real jets in binary systems.

The gravitational potential of the two stars with assumed masses of 1 
$\mbox{M}_{\odot}$ each is considered. However, the effect of the gravitational
potential on the resulting jet structure is marginal. 

The numerical resolution was chosen to 20 grid cells per AU, therefore our 
computational domain was 1000$\times$600 grid cells. To account for the 
counter-jet and the other part of the jet, respectively, the boundary 
conditions in the equatorial plane and on the jet axis are set to reflection 
symmetry. On the other boundaries, outflow conditions are chosen.

\subsection{Parameters for the pulsed jet}

The jet is produced within a thin jet nozzle with a radius of 1 AU. Because we 
like to investigate in these simulations the propagation of small scale 
structures in the jet and not the formation and collimation of the jet 
outflow, this ansatz is appropriate. We assume that the jet is already 
completely collimated when leaving the nozzle. To simulate the stable velocity 
component mentioned, the initial velocity of the jet is chosen to 1000 km 
s$^{-1}$ or 0.578 AU d$^{-1}$ and its density is set to 
$8.4 \cdot 10^{-18}$ g cm$^{-3}$ (equal to a hydrogen number density of 
$5 \cdot 10^6$ cm$^{-3}$). These parameters lead to a density contrast 
$\eta$ of $5 \cdot 10^{-3}$, a Mach number of $\approx 60$ in the nozzle and a 
mass loss rate of $\approx 10^{-8}$ $\mbox{M}_{\odot}$ yr$^{-1}$. 

Repeatedly each seventh day, the velocity and density values in the nozzle are 
changed to simulate the jet pulses which are seen in the observations
of MWC 560. The effects of different pulse densities and speeds are 
investigated with a parameter study for adiabatic jet models. 

\section{Jet structure} \label{sec_struc}

\subsection{Adiabatic models}

We present eight adiabatic models for pulsed underdense jets which differ in 
their pulse parameters. 

Table \ref{pulses} lists the main parameters of the models: the 
model number, the density of the pulse in the nozzle n$_{\rm{pulse}}$ in 
cm$^{-3}$, the velocity of the pulse in the nozzle v$_{\rm{pulse}}$ in cm 
s$^{-1}$, the mass loss during the pulse $\dot M$ in g s$^{-1}$ and in 
$\mbox{M}_{\odot}$ yr$^{-1}$ and the kinetic jet luminosity during
the pulse in erg s$^{-1}$. Each pulse lasts for one day. Thus the first 
4 models have pulses with enhanced velocity and lower, the same or higher 
density as the ``regular''outflow. In the models v to viii only the density is 
changed, except for the special model vii which has no pulses at all. Columns 
7 and 8 in Table 1 give the axial and radial extent of the bow shock after 380 
days of simulation time at which the first model (model iv) 
reaches the outer boundary. The bow shock sizes are results of our simulations 
which will be discussed below. 

\begin{table*}
\caption{Parameters of the jet pulses: the model number, the density of the 
pulse in the nozzle n$_{\rm{pulse}}$ in cm$^{-3}$, the velocity of the 
pulse in the nozzle v$_{\rm{pulse}}$ in cm s$^{-1}$, the mass outflow during 
the pulse $\dot M$ in g s$^{-1}$ and in $\mbox{M}_{\odot}$ yr$^{-1}$, the 
kinetic jet luminosity during the pulse in erg s$^{-1}$ and the axial and 
radial extent of the bow shock after 380 days of simulation time. Model vii is 
a special case, because no pulses are present. Therefore ``pulse'' values 
given for this model are identical to the jet parameters out of pulses valid 
for each model} \label{pulses}
\begin{flushleft}
\begin{tabular}{lcccccccc}
\hline
model & n$_{\rm{pulse}}$ [cm$^{-3}$] & v$_{\rm{pulse}}$ [cm s$^{-1}$] & 
$\dot M$ [g s$^{-1}$] & $\dot M$ [$\mbox{M}_{\odot}$ yr$^{-1}$] & 
L$_{jet}$ [erg s$^{-1}$] & 
$z_{\rm{b.\,s.}}$ [ AU ] & $r_{\rm{b.\,s.}}$ [ AU ] \\
\hline
i & $1.25 \cdot 10^6$ & $2.0 \cdot 10^8$ & $2.94 \cdot 10^{17}$ & 
$4.66 \cdot 10^{-9}$ & $5.88 \cdot 10^{33}$ & 
41.2 & 15.2 \\
ii & $2.5 \cdot 10^6$ & $2.0 \cdot 10^8$ & $5.88 \cdot 10^{17}$ & 
$9.33 \cdot 10^{-9}$ & $1.18 \cdot 10^{34}$ & 
42.0 & 15.4 \\
iii & $5.0 \cdot 10^6$ & $2.0 \cdot 10^8$ & $1.18 \cdot 10^{18}$ & 
$1.87 \cdot 10^{-8}$ & $2.35 \cdot 10^{34}$ & 
46.0 & 17.1 \\
iv & $1.0 \cdot 10^7$ & $2.0 \cdot 10^8$ & $2.35 \cdot 10^{18}$ & 
$3.73 \cdot 10^{-8}$ & $4.70 \cdot 10^{34}$ & 
50.0 & 18.4 \\
v & $1.25 \cdot 10^6$ & $1.0 \cdot 10^8$ & $1.47 \cdot 10^{17}$ & 
$2.33 \cdot 10^{-9}$ & $7.35 \cdot 10^{32}$ & 
43.8 & 14.4 \\
vi & $2.5 \cdot 10^6$ & $1.0 \cdot 10^8$ & $2.94 \cdot 10^{17}$ & 
$4.66 \cdot 10^{-9}$ & $1.47 \cdot 10^{33}$ & 
48.6 & 14.6 \\
vii$\, ^{*}$ & $5.0 \cdot 10^6$ & $1.0 \cdot 10^8$ & $5.88 
\cdot 10^{17}$ & $9.33 \cdot 10^{-9}$ & $2.93 \cdot 10^{33}$ & 
48.0 & 14.4 \\
viii & $1.0 \cdot 10^7$ & $1.0 \cdot 10^8$ & $1.18 \cdot 10^{18}$ & 
$1.87 \cdot 10^{-8}$ & $5.88 \cdot 10^{33}$ & 
38.7 & 14.5 \\
\hline
\end{tabular}
\\
\noindent
$^*$ equivalent to no pulses, these values represent the jet parameters 
out of pulses valid for each model
\end{flushleft}
\end{table*}

\begin{figure}
   \resizebox{\hsize}{!}{\includegraphics{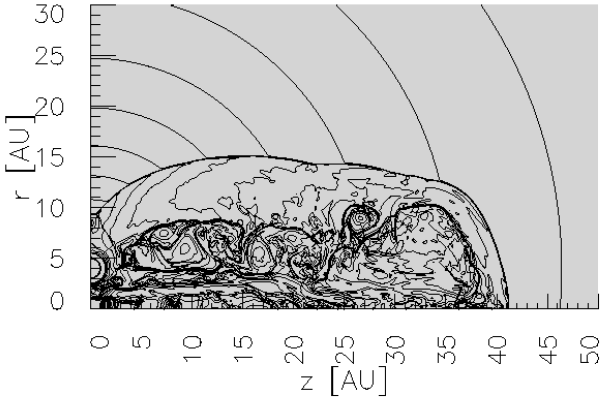}}   
   \caption{Contour plot of the logarithm of density at day 380; model i 
     ($v_{\rm{pulse}}=2000$ km s$^{-1}$, $n_{\rm{pulse}}=1.25 
     \cdot 10^6$ cm$^{-3}$)}
   \label{densityarray1}
\end{figure}

In Fig. \ref{densityarray1} the loga\-rithm of density is plotted 
for model i of the eight hydrodynamical models listed in Table \ref{pulses}. 
 The density plots of the other seven simulations are qualitatively similar 
and therefore omitted.

In the models i--iv with the higher jet pulse velocity of 2000 km 
s$^{-1}$, the axial extent of the bow shock after 380 days -- and therefore 
its averaged velocity -- is increasing with increasing jet pulse density and 
mass outflow (see Table \ref{pulses}). This seems to be a 
trend in the simulations which is also valid for the radial extent of the bow 
shock. No clear trend is present in the four models v--viii, where the jet 
velocity is constant and only the jet pulse density is varied. The models with 
the highest and lowest pulse density have a lower averaged bow shock velocity 
than the intermediate models. The radial extent of the bow shock, however, is 
for all four models v--viii practically equal.

Thus we can say that the shock front expansion velocity is about equal within 
$10-20$ per cent for all our adiabatic jet simulations. The average expansion 
velocity of the bow shock during the first year is about 
$200~{\rm km}\,{\rm s}^{-1}$ in axial direction and about 
$75~{\rm km}\,{\rm s}^{-1}$ in radial direction. This has to be compared with 
the gas velocity in the jet nozzle, which is $1000~{\rm km}\,{\rm s}^{-1}$.

\begin{figure}
   \resizebox{\hsize}{!}{\includegraphics{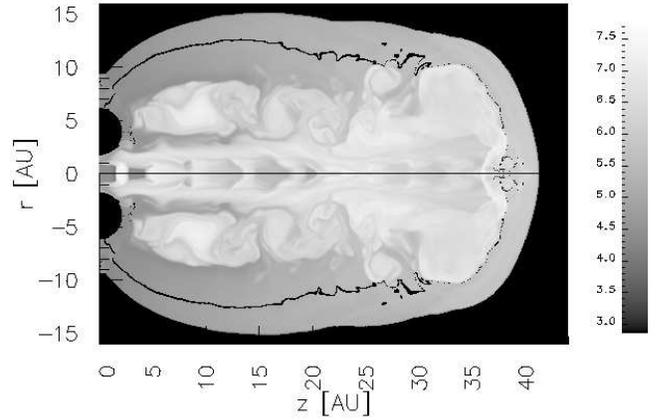}}
   \caption{Logarithm of temperature of the model i; temperatures are present 
     in the range of $10^3$ K in the red giant wind, of $10^5$ K in the 
     shocked ambient medium and up to $10^8$ K in the jet beam and the cocoon; 
     in this and the following color plots, only the jet material is 
     considered which is filtered out by means of a passively advected 
     tracer, the black line inside the jet cocoon is an artifact resulting 
     from this filtering}
   \label{NV_2.0_temp}
\end{figure}

In the models, temperatures are present in the range of $10^3$ K in the red 
giant wind and $10^5$ - $10^8$ K in the jet (Fig. \ref{NV_2.0_temp}). These are
by far too high for the observed absorptions from neutral or singly ionized 
metals which are only expected for cool gas with temperatures around 
$T\approx 10000$ K or lower. 

It is well known from previous simulations of protostellar and extragalactic
jets \citep[e.g.][and references therein]{StN,SGW,GPC} that the resulting 
jet structure differs strongly between purely hydrodynamical models and models 
using radiation hydrodynamics. This should also be the case in the parameter 
region of jets in symbiotic stars.

An estimate on the cooling time for the jet gas due to bremsstrahlung can be 
obtained from  
\begin{equation}
t_{cool} = \frac{n\,k\,T}{1.68 \cdot 10^{-27}\,T^{1/2}\,n_{e}\,n_{i}} = 0.83\,
\frac{\sqrt{T_{7}}}{n_{7}}\, \rm{yr}
\end{equation}
\noindent
where $T_{7}$ is the temperature in units of $10^7$ K which is about the 
postshock temperature of shocks with velocities of 1000 km s$^{-1}$ \citep{DoS}
and $n_{7}$ the number density in units of $10^7$ cm$^{-3}$, which are typical 
for our adiabatic models. It follows that the cooling time due to 
bremsstrahlung is comparable to the propagation time of the jet. This implies 
that radiative cooling is important for the employed jet model parameters and 
should be included in the energy equation. 

\subsection{Model with cooling}

We performed one pulsed jet simulations including radiative
cooling. For this we have chosen the same parameters as for the hydrodynamical 
model i, in which the jet velocity during the pulse phase is doubled from
1000~km\,s$^{-1}$ to 2000~km\,s$^{-1}$ while the particle number density is 
reduced by a factor of 4 from $5\cdot 10^6\,{\rm cm}^{-3}$ to 
$1.25\cdot 10^6\,{\rm cm}^{-3}$ in the jet beam. Thus the kinetic energy of the
jet gas in the nozzle remains constant. 

Radiative cooling is treated with a simple procedure. The cooling by metals is 
described by a general cooling function $\Lambda(T)$ adopted from \citet{SuD} 
and bremsstrahlung emission is considered. In addition we determined for each 
grid point the density of \ion{H}{i}, \ion{H}{ii} and e$^{-}$ from the rate 
equations and calculate the cooling due to collisional ionization and line 
excitation of \ion{H}{i}.

The expensive computational costs of the additional terms and equations to 
solve, together with an increased demand for memory, made it impossible to 
perform this simulation on a normal workstation. Therefore the already 
vectorized code was expanded to run on the NEC SX-5 supercomputer at the High 
Performance Computing Center (HLRS) in Stuttgart \citep{Kra}. 

\begin{figure}
   \resizebox{\hsize}{!}{\includegraphics{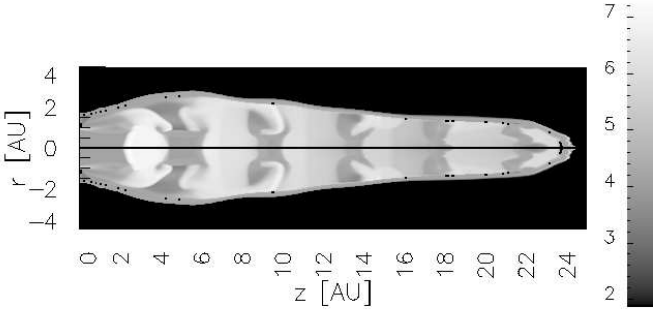}}
   \caption{Logarithm of temperature of the model i with cooling; again the 
     highest temperatures are present in the cocoon and jet beam, the region 
     of medium temperatures is shrinked with respect to model i without 
     cooling}
   \label{NV_3.0_temp}
\end{figure}

\begin{figure}
   \resizebox{\hsize}{!}{\includegraphics{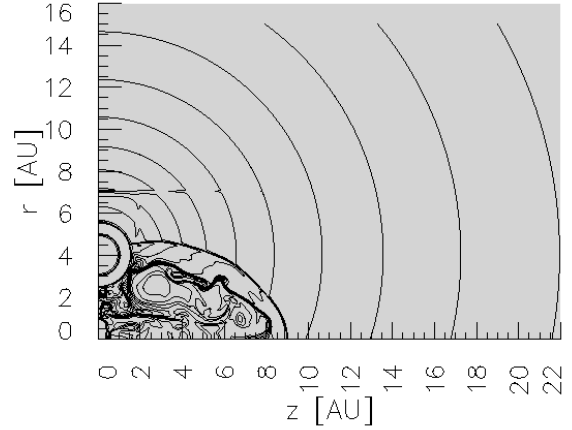}}   
   \caption{Contour plot of the logarithm of density at day 74; model i without
     cooling ($v_{\rm{pulse}}=2000$ km s$^{-1}$, 
     $n_{\rm{pulse}}=1.25 \cdot 10^6$ cm$^{-3}$)}
   \label{modi_det}
\end{figure}
\begin{figure}  
   \resizebox{\hsize}{!}{\includegraphics{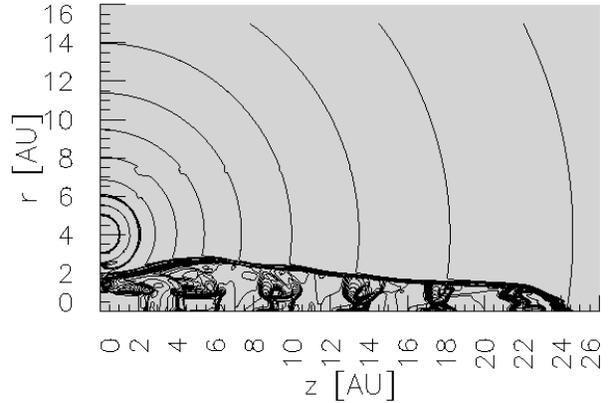}}  
   \caption{same as figure \ref{modi_det}; model i with cooling 
     ($v_{\rm{pulse}}=2000$ km s$^{-1}$, $n_{\rm{pulse}}=2.5 \cdot 10^6$ 
     cm$^{-3}$)}
   \label{modcool_det}
\end{figure}

The cooled jet looses energy through radiation which instantaneously lowers 
the pressure and the temperature (Fig. \ref{NV_3.0_temp}). Therefore the 
radial extension of the jet, the cross section and the resistance exerted by 
the external medium are weakened. This leads to a faster propagation velocity. 
In Figs. \ref{modi_det} and \ref{modcool_det}, the jet 
density structure after 74 days is plotted for model i without and with 
cooling, respectively. This illustrates well the dramatic difference in the jet
propagation.
The bow shock velocity shows an apparently steady increase 
from about $500~{\rm km}\,{\rm s}^{-1}$ during the first days to about 
$730~{\rm km}\,{\rm s}^{-1}$ after 70 days. This steadiness, 
however, could be pretended by our coarse time resolution of 1 day. The 
average bow shock velocity for the covered 74 days is 
$570~{\rm km}\,{\rm s}^{-1}$. The bow shock 
velocity is increasing with time, because the jet head area remains almost 
constant and is therefore not able to compensate the decreasing local density 
contrast as in the simulations without cooling. The maximum radial extend of 
the jet cocoon is about 2.5~AU. 

\subsection{Jet structure}

\begin{figure}
   \resizebox{\hsize}{!}{\includegraphics{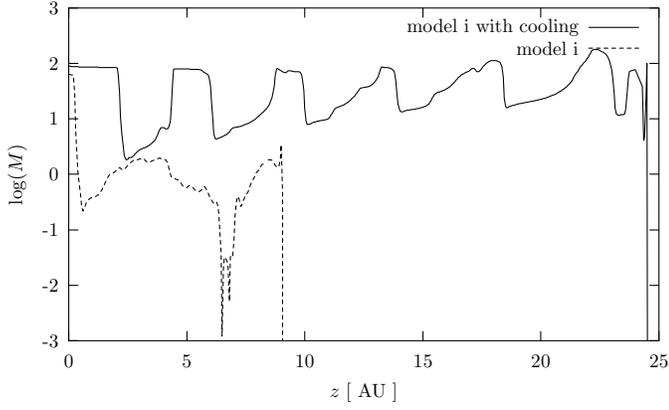}}
   \caption{Cut of the logarithm of Mach number along the jet axis for 
     model i with cooling and without cooling. Results for both models are for 
     day 74 in the simulation, for which 2D plots are plotted in Figs. 
     \ref{modi_det} and \ref{modcool_det} respectively.}
   \label{slice_cool_mach}
\end{figure}

\begin{figure}
   \resizebox{\hsize}{!}{\includegraphics{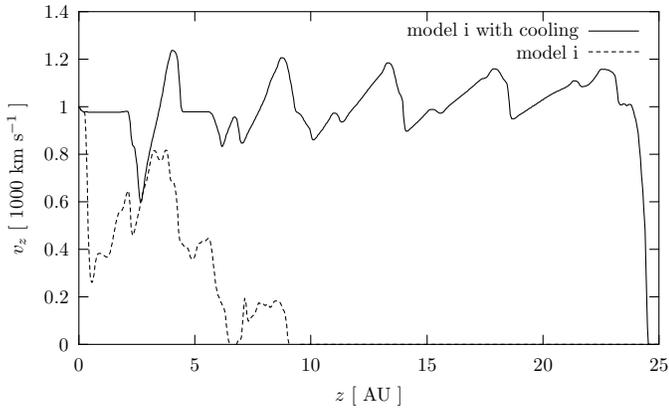}}
   \caption{Cut along the jet axis for model i as in Fig. 
     \ref{slice_cool_mach}, but of the parallel velocity component}
   \label{slice_cool_vx}
\end{figure}

\begin{figure}
   \resizebox{\hsize}{!}{\includegraphics{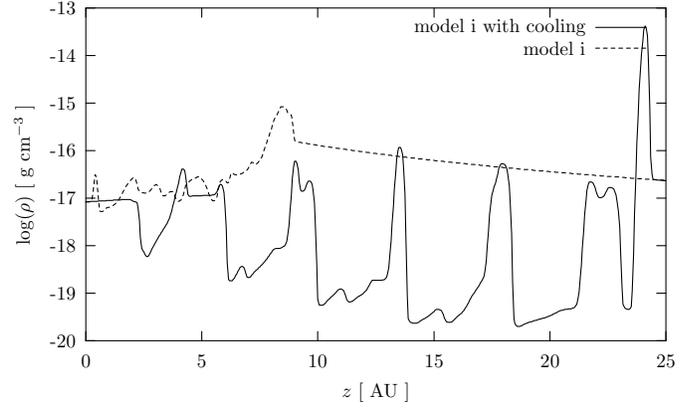}}
   \caption{Cut along the jet axis for model i as in Fig. 
     \ref{slice_cool_mach}, but of the logarithm of density}
   \label{slice_cool_den}
\end{figure}

\begin{figure}
   \resizebox{\hsize}{!}{\includegraphics{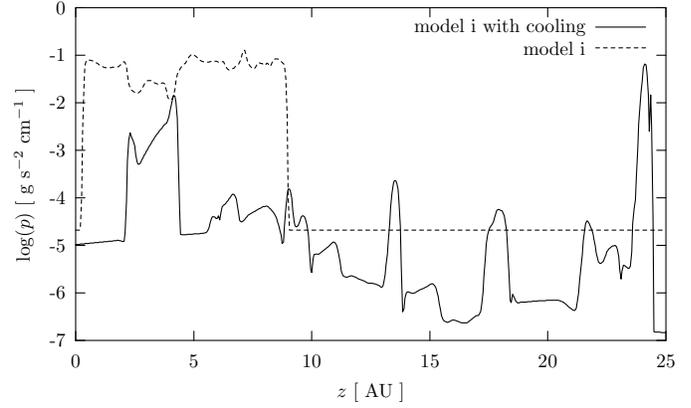}}
   \caption{Cut along the jet axis for model i as in Fig. 
     \ref{slice_cool_mach}, but of the logarithm of pressure}
   \label{slice_cool_p}
\end{figure}

\begin{figure}
   \resizebox{\hsize}{!}{\includegraphics{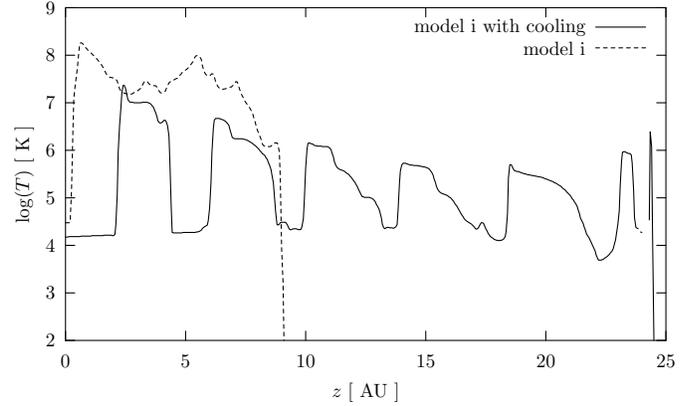}}
   \caption{Cut along the jet axis for model i as in Fig. 
     \ref{slice_cool_mach}, but of the logarithm of temperature}
   \label{slice_cool_T}
\end{figure}

From our simulations we can investigate in detail the internal structure of 
the model jets. In Figs. 
\ref{slice_cool_mach}-\ref{slice_cool_T} axial cuts along the jet of 
hydrodynamical quantities are plotted for model i without and with cooling. 
However, in the following discussion we focus on the calculations with 
radiative cooling. The cooled jet shows a very simple, periodic jet structure 
for the Mach number, axial velocity, density, pressure and 
temperature (Figs. \ref{slice_cool_mach}-\ref{slice_cool_T}). The internal 
shocks, which have not yet merged with the bow shocks, can be identified with 
well defined discontinuities in all parameters. 

\begin{figure}
   \resizebox{\hsize}{!}{\includegraphics{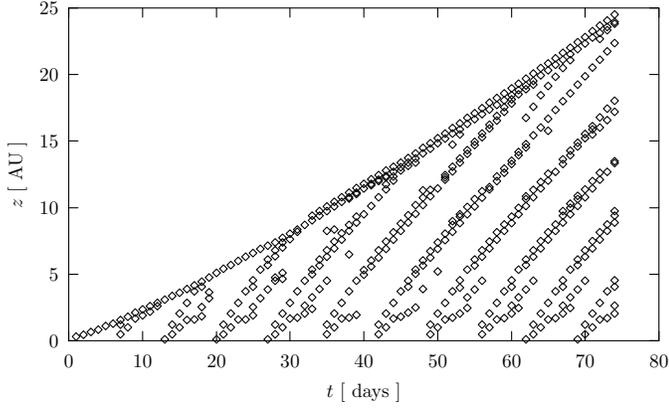}}
   \caption{Evolution of the pulses (model i with cooling); all pulses can be 
     traced directly without the presence of the KH-instabilities seen in 
     model i without cooling; the diamonds represent the
     locations of the pulses derived from the slice of the Mach number by 
     searching for extremal points in Fig. \ref{slice_cool_mach}, the splitting
     of some knots is due the inner structure of their peaks}
   \label{NV_3.0_shock}
\end{figure}

The jet flow shows thus two very different states of gas parameters
which we call knots and hot beam. In the knots the density is high $\approx
10^{-16}\,{\rm g}\,{\rm cm}^{-3}$ and the temperature low $\approx 10^4$~K.
Contrary to this the density in the hot beam is about 1000 times lower or   
$\approx 10^{-19}\,{\rm g}\,{\rm cm}^{-3}$ while the temperature is much 
higher between $10^{5.5}$~K to $10^7$~K.
Fig. \ref{NV_3.0_shock} shows the evolution of the internal shocks along the 
jet axis. The locations of the pulses were derived from the 
slice of the Mach number by searching for extremal points in Fig. 
\ref{slice_cool_mach}. The splitting of some knots is due the inner structure 
of their peaks. The propagation of all pulses can be traced in the 
plot. This is in good accordance with simple theoretical models of pulsed jets 
\citep{RaC}.  
Each new pulse is instantaneously slowed down from 
$2000~{\rm km}\,{\rm s}^{-1}$ to about $1200~{\rm km}\,{\rm s}^{-1}$
within the first two days. Without the disturbance of the KH-instabilities, 
the distance between the internal shocks created by the periodic velocity 
pulses stays constant. It remains to be investigated whether this situation 
changes if the kinetic energy for the jet gas in the pulses and the ``normal'' 
beam in the nozzle are not the same as in this model.   

\subsection{Emission plots}

We present emission plots of the cooled jet model in bremsstrahlung, 
synchrotron and optical radiation. 

In the models, temperatures are present in the range of $10^3$ K in the red 
giant wind and $10^4$ - $10^7$ K in the jet (Figs. \ref{NV_3.0_temp}
and \ref{slice_cool_T}). The high gas 
temperature ($T > 10^5$ K) in the low density region of the jet beam
makes the jets X-ray emitters due to thermal bremsstrahlung. 

According to standard radiation theory, e.g. \citet{RyL}, the total 
emissivity due to bremsstrahlung due to a completely ionized plasma is
\begin{equation}
j = 1.68 \cdot 10^{-27}\,T^{1/2}\,n_{e}\,n_{i}\, \rm{erg\,s}^{-1} 
\rm{cm}^{-3} .
\end{equation}

Using the density and pressure data in each grid cell of the simulations, 
the emission per grid cell can be calculated and emission plots can be 
produced.

Synchrotron emission should be present due to the acceleration of electrons in 
the shocks and the presence of local magnetic fields in the plasma. With
a power-law energy distribution of the electron $N(E)\,dE = E^{-\kappa}\,dE$, 
the spectral index of the resulting spectrum of the emission is 
$\alpha = (\kappa - 1)/2$ \citep{RyL} and the total emissivity of synchrotron 
radiation can be estimated to
\begin{equation}
j = \epsilon_{s}\,p\,B^{1+\alpha} = \epsilon_{s}\,\beta
\,p^{(3+\alpha)/2}
\end{equation}

\noindent
with an assumed spectral index $\alpha = 0.6$ \citep{SSB}, $\epsilon_{s}$ the
normalization of the power-law and $\beta < 1$ the proportionality 
factor of equipartition. This is a very crude estimate, furthermore the values 
of the parameters are quite unclear. Therefore the synchrotron emission plots 
should be considered only as relative and qualitative.

According to \citet{All}, we can choose those grid cells to estimate 
the \ion{H}{i} line emission (of only Balmer lines), in 
which the temperature is 
larger than $10^4$ K and where only recombination contributes to the emission. 
Hydrogen is assumed to be fully ionized. The emissivity is then 
\begin{equation}
j = 4.16 \cdot 10^{-25} \, T_{4}^{-0.983}
\,10^{-0.0424/T_{4}}\,n_{e}\,n_{i}\, \rm{erg\,s}^{-1} \rm{cm}^{-3}
\end{equation}

\noindent
with $T_{4}$ the temperature in units of $10^4$ K.

\begin{figure}
   \resizebox{\hsize}{!}{\includegraphics{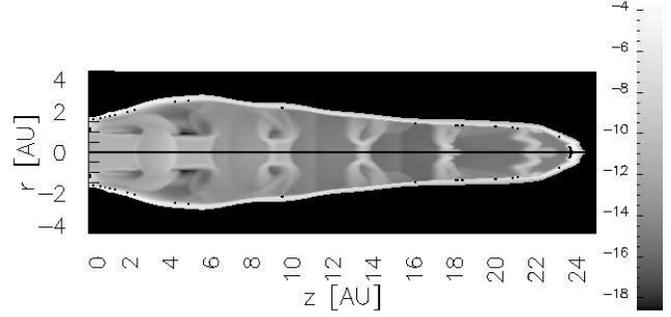}}
   \caption{Logarithm of Bremsstrahlung emission of the model i with cooling 
     in $\rm{erg\,s}^{-1} \rm{cm}^{-3}$; the shocked ambient medium 
     collapsed, therefore the emission features of the jet beam are visible in 
     the whole jet width}
   \label{NV_3.0_brems}
\end{figure}
\begin{figure}
   \resizebox{\hsize}{!}{\includegraphics{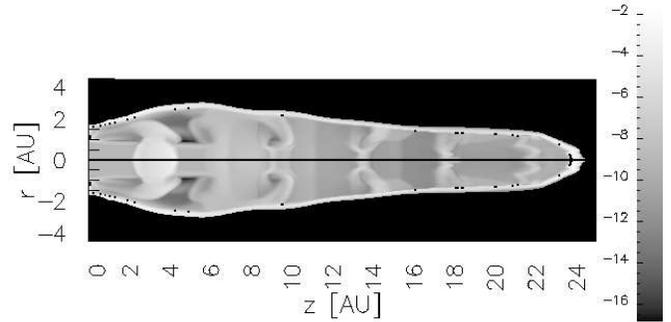}}
   \caption{Logarithm of Synchrotron emission of the model i with cooling in 
     $\rm{erg\,s}^{-1} \rm{cm}^{-3}$}
   \label{NV_3.0_synch}
\end{figure}
\begin{figure}
   \resizebox{\hsize}{!}{\includegraphics{NV_3.0_opt.epsf}}
   \caption{Logarithm of \ion{H}{i} line emission of the 
     model i with cooling in $\rm{erg\,s}^{-1} \rm{cm}^{-3}$}
   \label{NV_3.0_opt}
\end{figure}
\noindent
The integrated luminosities at day 74 are: 
\begin{eqnarray}
L_{\rm{Bremsstrahlung}} &=& 1.15 \cdot 10^{33}\, \rm{erg\,s}^{-1} 
\nonumber \\
L_{\rm{HI}} &=& 6.96 \cdot 10^{32}\, \rm{erg\,s}^{-1} \nonumber
\end{eqnarray}

The integrated synchrotron emission was omitted because it is only a relative 
and qualitative presentation. $L_{\rm{HI}}$ accounts only for the \ion{H}{i} 
recombination emission. The total cooling emission in all emission lines is 
certainly higher. Thus the dominant cooling radiation should be atomic 
emission lines. 

For the comparison of the emission plot with observations it should be
considered that the \ion{H}{i} plot would not be representative for all 
optical lines. For example emission from collisionally excited lines like 
\ion{Ca}{ii} is only expected from the very coolest $T< 15000$~K regions in 
the jet knots. No emission would come from the hot jet beam or the cocoon 
because there Ca would be in a higher ionization state than \element[+]{Ca}. 
Thus the knot structure may be particularly well defined in such low 
ionization lines. Similar selection effects must be considered for the 
observations of the Bremsstrahlung radiation. Thermal free-free radiation in 
the radio range would strongly favor the low temperature emission regions, 
such as the dense knots, while the Bremsstrahlung radiation observable with 
X-ray telescope would only originate from the very hottest, low density 
regions in the hot beam or the cocoon. The plot for the Bremsstrahlung simply 
includes the emitted radiation in all wavelength bands. 
\begin{figure}
   \resizebox{\hsize}{!}{\includegraphics{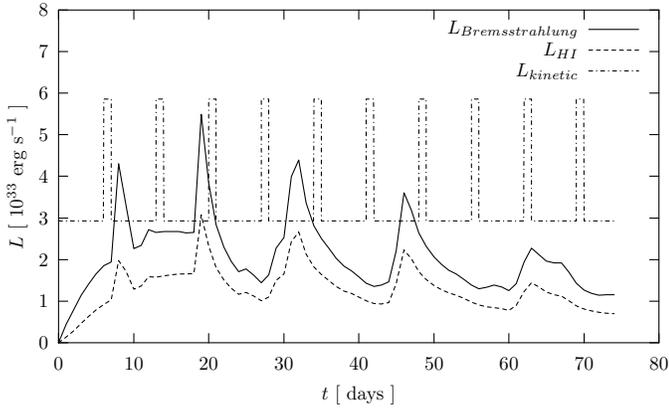}}
   \caption{Evolution of the kinetic, bremsstrahlung and \ion{H}{i} luminosity}
   \label{Lvst}
\end{figure}
Fig. \ref{Lvst} shows the evolution of the kinetic, bremsstrahlung and 
\ion{H}{i} luminosity until day 74. The emitted luminosity is about 50 per cent
of the kinetic luminosity. The peaks in the emitted luminosity coincide with 
the merging of the pulses with the bow shock (see Fig. \ref{NV_3.0_shock}).
These ``flashes'' are in accordance with simple analytical models \citep{RaC}.

\section{Calculations of jet absorption profiles} \label{sec_line_param}

Using the model results from the hydrodynamical simulations, we calculate 
the absorption line structure.

As emission region we define a disc in the equatorial plane ($z = 0$) 
which extends to a radius $r_{\rm em}$. For each grid cell inside this region, 
we assume a normalized continuum emission and a Gaussian emission line profile:
\begin{equation} \label{initial_line}
I_{0} = 1 + I_{\rm peak} \cdot \exp{\left( - v^2/\sigma^2 \right)} .
\end{equation}
\noindent
This initial intensity is taken to be independent of position $r < r_{\rm em}$.

Further we consider the possibility of different system inclinations $i$ for 
the calculation of light rays from the emission region through the 
computational domain. We consider only absorption along straight lines and 
neglect possible emissivity or scattering inside the jet region. 

The initial continuum and line emission (\ref{initial_line}) is taken as input 
for the innermost grid cell of each path. We then calculate the absorption 
within each grid cell $j$ along the line of sight according to:
\begin{eqnarray} \label{eq_abs_cs}
I_{j} &=& I_{j-1} \cdot e^{-\tau_\lambda}  \nonumber \\
 \tau_\lambda &=& \frac{\pi \, e^2}{m_{e} \, c} \, 
 \lambda_{kl} \, \left( 1 - \frac{v_{j}}{c} \right) \, 
 \frac{\rho_{j}\,\eta}{m_{H}} \, \Delta x_{j}\,f_{kl}  \nonumber \\
 &=& \mathcal{C}_{kl}\,\Delta x_{j}\,\mathcal{F}(v_{j},\rho_{j})\,,
\end{eqnarray} 
\noindent
where $e$ is the electron charge, $m_{e}$ the electron mass, $c$ the speed of 
light, $\lambda_{kl}$ the rest frame wavelength of the transition, $m_{H}$ the 
proton mass and $f_{kl}$ the oscillator strength. The parameters are constant 
for a given atomic transition. $\Delta x_{j}$ is the length of the path 
through grid cell $j$ which is a function of the inclination $i$. 

The parameters depending on the different hydrodynamical models are the 
velocity projected onto the line of sight $v_{j}$, the mass density $\rho_{j}$ 
and $\eta$ the relative number density with respect to hydrogen 
$\eta=n_k/n_{\rm H}$ of the absorbing atom in the lower level $k$ of the 
investigated line transition. The velocity is binned for the calculation of 
the absorption. The size of the bins $\Delta v$ can be interpreted as a 
measure of the kinetic motion and turbulence in one grid cell. This absorption 
dispersion helps to smooth the effects of the limited spatial resolution of 
the numerical models.

The absorption calculation through the jet region is repeated for all 
possible light paths from the emission region to the observer. The arithmetic 
mean of the individual absorption line profiles from all path is then taken as 
resulting spectrum.

\subsection{Model parameters}

The quantities determining the absorption line profiles are first the
parameters defining the hydrodynamical model. These are for our model grid the 
pulse velocity, the pulse density, and the time (day) in the simulation. For 
model i we have to distinguish further between adiabatic calculation and
calculations including radiative cooling. 

For the emission from the jet source, the size of the emission region 
$r_{\rm em}$ must be fixed. For the emission spectrum we adopt throughout this 
paper a normalized continuum and an emission line component at rest velocity 
(the same transition as the calculated jet absorption line) with 
$I_{\rm peak}=6$, $\sigma=100$ km s$^{-1}$. An additional parameter is the 
inclination $i$ under which the system is seen.

The parameters of the absorbing transition are defined by the rest wavelength
$\lambda_{kl}$, the oscillator strength $f_{kl}$ and the population of the 
absorbing atomic level $\eta=n_k/n_{\rm H}$. Further we have to define the 
velocity bin size $\Delta v$ which is a measure for the adopted turbulent (and 
kinetic) motion of the absorbers. 

\subsection{Results for different model parameters}

\begin{figure}
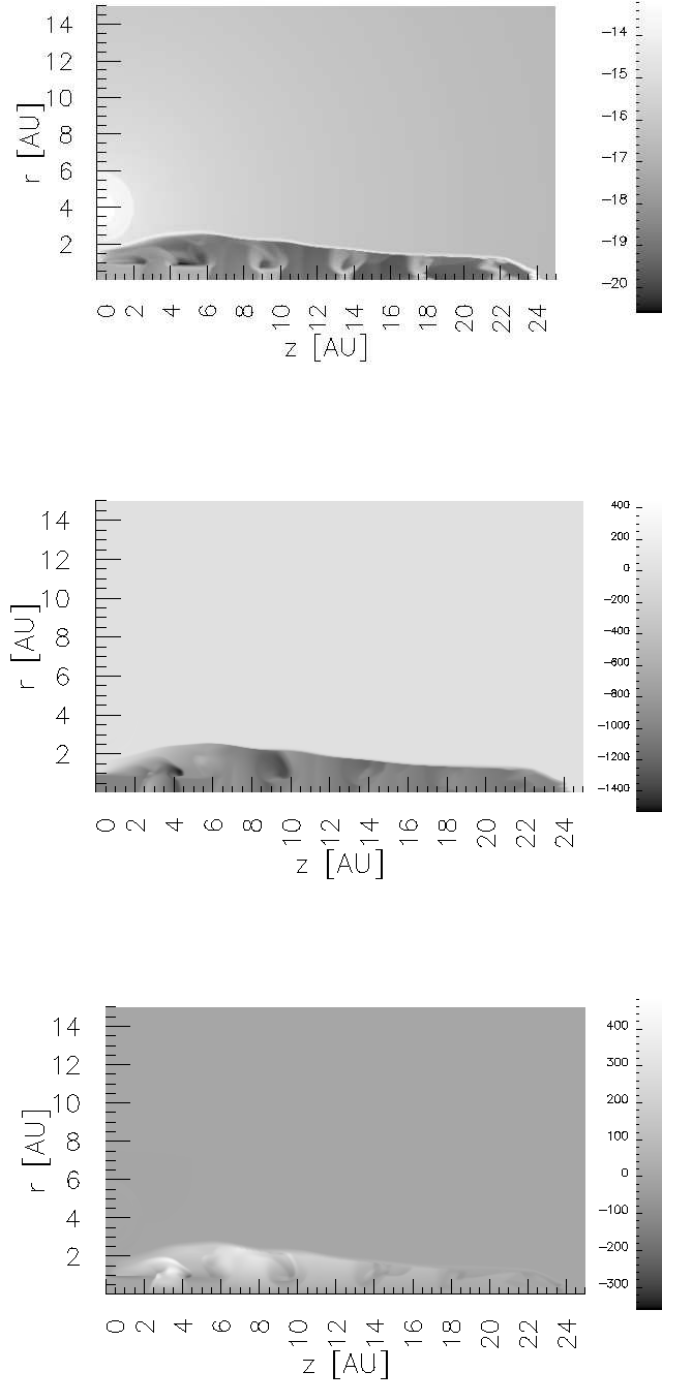

   \resizebox{\hsize}{!}{\includegraphics{NV3.0_074_den.epsf}}
   \resizebox{\hsize}{!}{\includegraphics{NV3.0_074_vx.epsf}}
   \resizebox{\hsize}{!}{\includegraphics{NV3.0_074_vy.epsf}}
   \caption{Pulsed jet model i with radiative cooling at day 74: Grey scale
     plots of the logarithm of density in g cm$^{-3}$ (top), of 
     the velocity components parallel (middle) and perpendicular
     (bottom) to the jet axis in km s$^{-1}$}
   \label{modelicool_greyscale}
\end{figure}

The structure of the synthetic absorption profiles reflect in some way the 
hydrodynamical structure of the jet outflow. The hydrodynamical variables 
determining the absorption profile are the density and the velocity projected 
on to the line of sight. The density, the velocity component parallel and the 
velocity component perpendicular to the jet axis are plotted in 
Fig. \ref{modelicool_greyscale} for the model with radiative cooling at day 74.
While the adiabatic jet produces a very extended and highly structured jet 
cocoon, the model i with radiative cooling produces an almost ``naked'' jet 
beam with practically no jet cocoon, but with a well defined and sharp 
transition towards the ambient medium. 
\noindent
In the adiabatic model four different
velocity regions can be distinguished (Fig. \ref{fourregions}):
\begin{itemize}
\item[I] the jet beam with high outflow velocities of $-800$ to $-1400$ km
  s$^{-1}$, 
\item[II] the cocoon and bow shock region with velocities of $-10$ to $-800$ 
  km s$^{-1}$,
\item[III] the external medium, the companion star and its wind with 
  velocities around 0 km s$^{-1}$, and 
\item[IV] the backflow near and besides the jet head with velocities of 0 to 
  $+500$ km s$^{-1}$. 
\end{itemize}

\noindent
In the jet models with radiative cooling there exists neither an extended 
cocoon and bow-shock region nor a backflow region. These regions are 
suppressed into a narrow transition region between jet beam and external 
medium  due to the high density and the efficient cooling. Thus the model with
cooling has essentially only two velocity regions: 

\begin{itemize}
\item[I] the jet beam, and 
\item[III] the external medium.
\end{itemize}

\subsection{Synthetic absorption line profiles}

From our model simulations we can calculate from the density and velocity 
structure of the jet (e.g. Fig. \ref{modelicool_greyscale}) the synthetic 
absorption line profile for different line of sights.

For the emission region we assumed that it has an radius of $r_{\rm em}=1$ AU 
and the resulting spectrum is the mean of the corresponding line of sights 
distributed over this emission area. For the atomic transition we use the 
parameters $\lambda_{kl}=3934$ \AA, $f_{kl}=0.69$ and $\eta=2 \cdot 10^{-6}$ 
appropriate for the \ion{Ca}{ii} K transition. For $\eta$ we just assumed that 
all Ca-atoms are in the ground state of \element[+]{Ca}. This is useful for 
the investigation of the expected absorption line structures introduced by the 
different velocity regions. However, the gas temperature in the adiabatic 
models is far too high to have Ca in the form of \element[+]{Ca}. This will be 
discussed in Sect. \ref{sec_ionization}. The velocity bin width for the 
calculation of the absorption profiles is set to $\Delta v =10$ km s$^ {-1}$. 
The parameters for the emission line component in the initial spectrum are 
$I_{\rm peak} = 6$ and $\sigma = 100$ km s$^ {-1}$.
\begin{figure}
   \resizebox{\hsize}{!}{\includegraphics{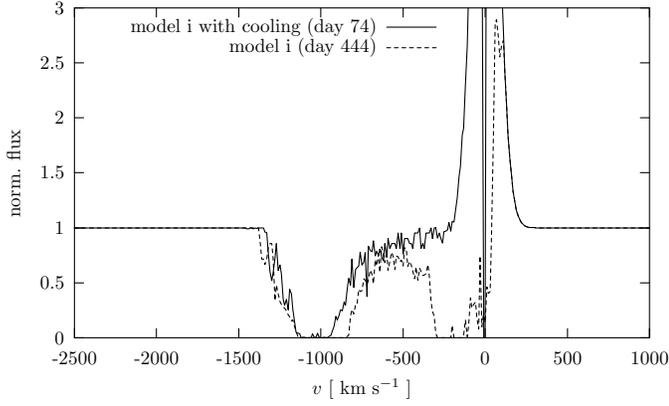}}
   \caption{Four different absorption regions in the adiabatic model i caused 
   by the jet beam, the bow shock region, the external medium and the backflow;
   in model i with cooling only the jet beam and the external medium are 
   clearly visible, while the bow shock region creates only a slight absorption
   and no backflow is present}
   \label{fourregions}
\end{figure}
Figure \ref{fourregions} shows the resulting synthetic absorption line profiles
calculated for the model i without cooling (adiabatic model) and with cooling. 

The model with cooling shows a strong, broad absorption from the velocity 
region I centered around RV $=-1100$ km$^{-1}$ from the jet beam. In addition 
there is a saturated, narrow absorption component in the middle of the emission
component caused by the almost stationary stellar wind of the cool giant 
(velocity region III) in front of the jet bow shock.

The absorption spectrum of the adiabatic model i is much more structured. 
Besides the velocity component I and III from the jet beam and the ambient 
medium respectively, there is also a broad velocity component from the bow 
shock region in the velocity range $v=-400$ to $+100$ km s$^{-1}$. This 
component can be associated with the RV-region II of the jet cocoon. The line 
of sight goes for this geometry not through the extended backflow region of 
the adiabatic jet. Signatures of the backflow regions may be visible for other 
line of sight geometries. 

Surprisingly, the structure of the main absorption trough is 
quite similar in both the adiabatic model and the model with radiative 
cooling.

\subsection{Different line of sight geometries}

In this section we describe the dependence of the jet absorption structure on 
the line of sight inclination and the emission region size for
the jet with radiative cooling. 

\begin{figure}
   \resizebox{\hsize}{!}{\includegraphics{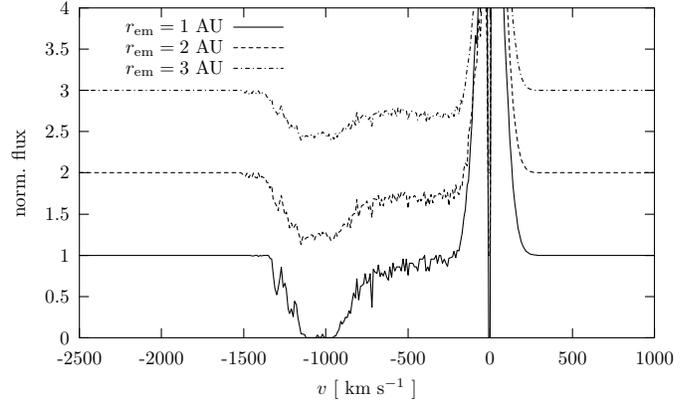}}
   \caption{Synthetic absorption line profile for three different radii  
     $r_{\rm em}=1,2,3$ AU of the emission region. Calculations for
     model i with cooling (Fig. \ref{modelicool_greyscale}) and inclination 
     $i = 0^{\circ}$. The appearance of absorption component created in the 
     jet beam and averaging effects are detectable}
   \label{line_rem}
\end{figure}

In Fig. \ref{line_rem} the synthetic absorption profiles are given for
different radii $r_{\rm em}=1, 2, 3$ AU of the disk-like emission region. The 
system inclination is set to $i=0^{\circ}$, therefore the projected velocity 
is equal to the velocity component parallel to the jet axis. 

For $r_{\rm em}=1$ AU the gas inside the jet beam (region I) with 
$v=-800$ to $-1300$ km s$^{-1}$ produces absorptions. The external medium 
(III), in front of the bow shock, absorbs the peak of the 
emission line. 
For higher values of $r_{\rm em}$ no additional structure appears, but the 
high velocity component is averaged out and therefore less strong.
\begin{figure}
   \resizebox{\hsize}{!}{\includegraphics{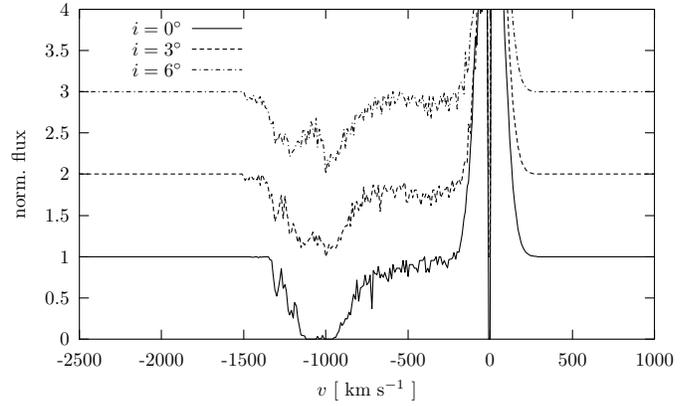}}
   \caption{Synthetic absorption line profile for the model i with cooling 
     seen under the inclination $i=0^{\circ}$, 3$^{\circ}$ and 6$^{\circ}$. 
     The radius of the emission region is $r_{\rm em} = 1$ AU. The splitting 
     of the main absorption component into two distinct components with 
     increasing system inclinations is visible.}
    \label{line_i}
\end{figure}
The absorption line structure depends also on the system inclination
as shown in Fig. \ref{line_i}. In this figure the synthetic spectra 
are compared for the inclinations $i=0^\circ, 3^\circ$ and $6^\circ$ for an 
emission region of radius $r_{\rm em} = 1$ AU. For $i=3^\circ$ and $i=6^\circ$ 
the absorption components are redistributed in comparison to the 
$i = 0^{\circ}$-case, because then other regions are along the line of sight 
while regions in the jet beam at large $z$ are missed. A region of weakened 
absorption appears and splits the main absorption component into two distinct 
components. 

\subsection{The velocity bin size}

\begin{figure}
   \resizebox{\hsize}{!}{\includegraphics{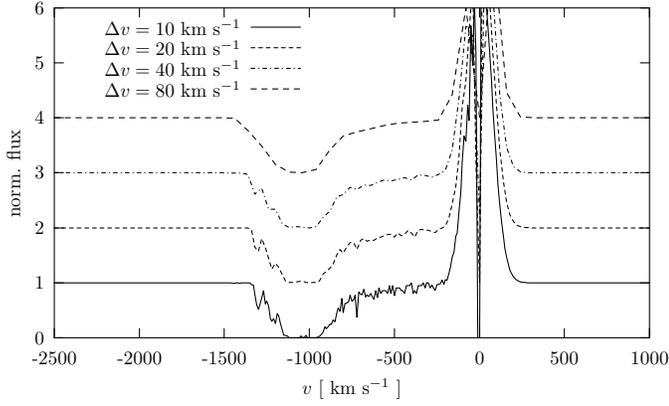}}
   \caption{Theoretical absorption line profile for values of the velocity 
     bin size $\Delta v$ (10, 20, 40, 80 km s$^{-1}$); higher velocity 
     dispersions lead to higher equivalent widths and smoother
     spectral absorption structures.}
   \label{line_dv}
\end{figure}

The velocity structure in the synthetic jet absorption profile represents an
average of the jet which is limited by the spatial resolution. The real 
velocity structure inside one grid cell is certainly more complex, mainly due 
to turbulent small scale gas motions. Therefore the discrete velocity value 
for the absorption of one grid cell should be replaced by a velocity 
dispersion. The strength of this dispersion, however, is not known. Thermal 
motion in our temperature regime leads to values of about 10 km s$^{-1}$, 
turbulence can increase them to 50 km s$^{-1}$ and more. The effect of such a
velocity dispersion can be considered in the calculation of the absorption 
line by the binning of the velocities into broader bins. For broader velocity 
bins, simulating a larger velocity dispersion of the absorbing atoms, a 
smoother jet absorption profile with less structure is obtained as shown in 
Fig. \ref{line_dv}. 
The observations show typically a smooth 
absorption line structure indicating a significant velocity turbulence.

\subsection{The ionization problem} \label{sec_ionization}

In all the jet absorption profile calculations in the previous sections we 
have assumed that all Ca atoms are singly ionized. 

In the adiabatic models the gas temperature is far 
too high for \element[+]{Ca}-ions. 
Assuming collisional ionization equilibrium, \element[+]{Ca} would be the 
dominant ionization stage for $T_e<15 000$ K. The relative abundance 
\element[+]{Ca}/Ca is less than 1 \% for $T_e>25 000$ K or less than 0.01 \% 
for $T_e>250 000$ K. After fitting the ionization balances from \citet{SuD} 
with
\begin{equation}
\eta ( T ) = \eta (0) \, \frac{1}{\exp(T/1000-14)+1}
\end{equation}
\noindent
and introducing this into the calculation, for our adiabatic model no 
absorption is present if ionization equilibrium is 
considered for the \element[+]{Ca} abundance. 

\begin{figure}
   \resizebox{\hsize}{!}{\includegraphics{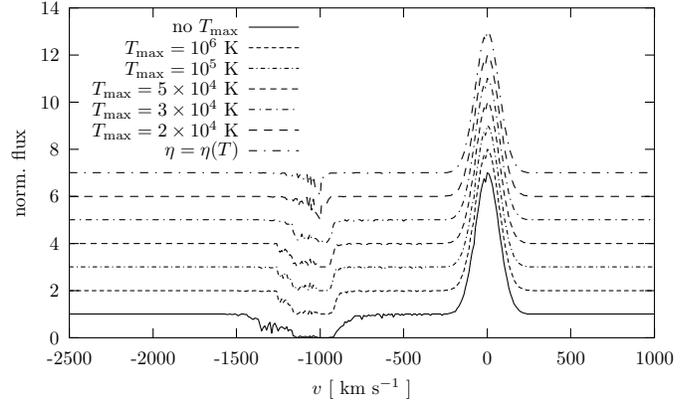}}
   \caption{Theoretical absorption line profile with and without a temperature 
     dependent $\eta$ for the model i with cooling (day 65)}
   \label{line_eta_T_cool}
\end{figure}

Even in the model with cooling, the jet gas 
temperature is slightly too high to produce a strong \ion{Ca}{ii} absorption 
(Fig. \ref{line_eta_T_cool}). The absorption 
generated by material which is cooler than 10$^5$ K is of the order of 60 \% 
in the model with cooling, the material with 
temperatues below $3 \times 10^4$ K is responsible for 48 \% of 
the absorption. Therefore 
the temperature in our model with cooling is only slightly overestimated. 
The high temperature of the model jet gas, compared 
to the observations, could be explained by unsufficient cooling. As the
cooling is proportional to the density squared, higher density gas in
the jet model would cool faster and down to the required temperature to 
account for the observed \ion{Ca}{ii} absorption. This may be achieved
with models having higher gas densities in the jet or pulse. Another 
possibility is, that high density clumps may form, for which the
cooling would be very efficient. However, the spatial resolution of
our model grid is too coarse for simulating such small scale structures.
A third reason could be the choice of the used cooling curves.
They extend only down to temperatures of $10^4$ K. Below that value, no cooling
is achieved in the present treatment. The adding of more cooling processes 
could improve the models and perhaps also solve the ionization problem.

\section{Time variability} \label{sec_time_var}

As next step the temporal evolution of the highest velocity component is 
investigated. This ``edge'' velocity for the variable absorption originates 
from the most recent pulse, close to the jet nozzle, which is not yet slowed 
down as much as the previous pulses. As this gas is located close to the 
continuum emission region, the corresponding absorption feature is present for 
all viewing angles considered and essentially independently from the
size of the emission region.

\begin{figure}
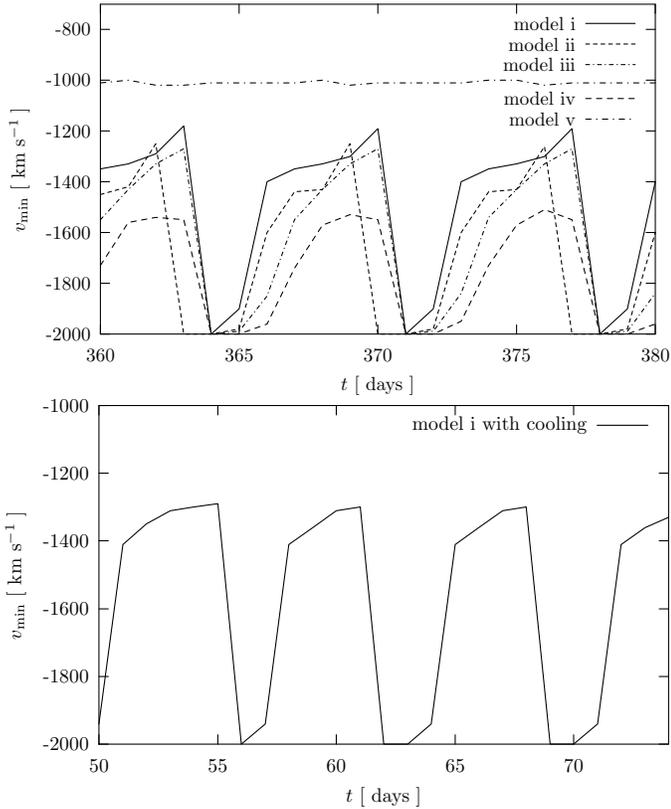

   \resizebox{\hsize}{!}{\includegraphics{time_var_i.eps}}
   \resizebox{\hsize}{!}{\includegraphics{time_var_cool.eps}}
   \caption{Time variability of the high (negative) velocity component for 
     models i-v (top) and for model i with cooling (bottom)}
   \label{time_var}
\end{figure}

In Fig. \ref{time_var}, the maximum outflow (negative) velocity of the high 
velocity component is plotted between day 360 and 380 for the first four 
adiabatic models and between day 50 and 74 for the one with cooling.
The first result is the fact, that pulses without velocity enhancements (only 
density changes) as investigated in the four adiabatic models v to viii, 
produce hardly any velocity effects. The velocity variations created by 
interactions between different density regions in the jet are only of the 
order of a few 10 km s$^{-1}$ and only in the model v, where the density in 
the pulse is the smallest. It seems unlikely that high velocity components can 
be produced by density jumps in the jet outflow. 

The simulations with high velocity pulses show all a qualitatively similar 
behavior, with a sudden outflow (negative) velocity increase to the pulse 
velocity of $-2000$ km/s followed by a velocity decay during the following 
days. Closer inspection of the evolution of the high velocity component shows 
some differences between the four adiabatic models i -- iv to the model i with 
cooling. In the latter, the velocity drops within one or two days from the 
initial --2000 km s$^{-1}$ to --1400 km s$^{-1}$ and then much slower --- 
apparently asymptotically --- to a stationary value of --1300 km s$^{-1}$ 
during the next four days. In the adiabatic model i, the first drop is similar,
but the slow decay seems to be only a plateau of two days, after that the 
velocity drops to --1150 km s$^{-1}$. The model ii shows the same behavior but 
with higher velocities (--1400 and --1250 km s$^{-1}$). The last two models 
iii and iv decay again in an asymptotic manner to --1300 km s$^{-1}$ and 
--1550 km s$^{-1}$, respectively.

The fact, that the structure and time variability of the high 
velocity components are quite similar in the adiabatic models and the model 
with radiative cooling, shows that also the former have their merits. As their 
computational requirements are by far smaller than those for models with 
radiative cooling, they can be used for (additional) parameter studies without 
a great loss of accuracy. Another point of future investigation could be then 
the influence of different jet pulse shapes and frequencies on the structure 
of the high velocity components.

\section{Jet absorption variability: comparison with observations} 

Based on our simulations of jet pulses we construct now sequences of absorption
line profiles which can be compared to the spectroscopic observations of the 
MWC 560 jet outflow. For the line of sight we adopt a direction parallel to 
the jet (inclination $i=0^\circ$) and an emission region size of 
$r_{\rm em}=1$ AU. Atomic parameters are taken for the \ion{Ca}{ii} absorption 
as in previous sections. 

\subsection{Variations of the jet absorption for adiabatic models}

For the comparison with the observations we consider only the adiabatic models
i - iv. One can directly see that the adiabatic models v - viii can be 
discarded because they do not show high velocity components in the synthetic 
absorption line profiles (Fig. \ref{time_var}, middle). For the models i to iv 
the density in the jet pulses increases from 0.25, 0.5, 1 to 2 times the 
interpulse jet density at the nozzle.

A major discrepancy in the jet absorption between our adiabatic simulations and
the observations is seen in the velocity region between $-300 - 0$ km s$^{-1}$ 
where the calculated absorption is grossly overestimated.

\begin{figure}
   \resizebox{\hsize}{!}{\includegraphics{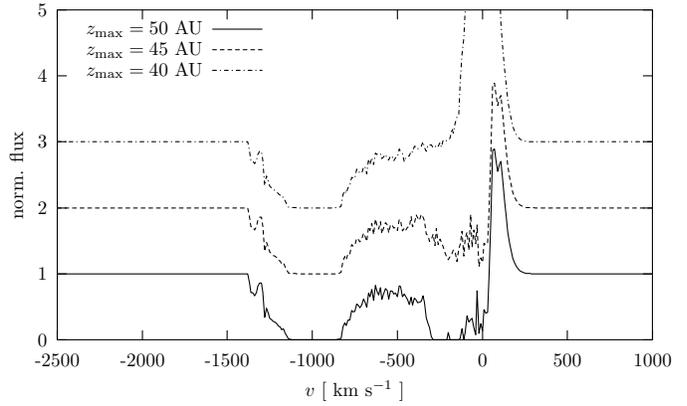}}
   \caption{Theoretical absorption line profile of the adiabatic model i for 
     values of $z_{\rm max}$; the absorption feature at RV $\approx$ 
     -300 km s$^{-1}$ is produced by the bow shock-Mach disk region}
   \label{line_zmax}
\end{figure}

This feature appears whenever the light path from the emission region to the 
observer travels through the jet head (bow shock-Mach disk) region (Fig. 
\ref{line_zmax}). The simulations were stopped when the jet reached the 
opposite boundary. At this position, the density of the external medium has 
dropped to the density value of the jet at its nozzle. In reality, the jet 
would extend further into a region where the density of the external medium is 
by far smaller. Then the densities in the jet head region also should be 
smaller and therefore the depth of the absorption. To account for this, we 
disregard the absorption produced in this region. Thus the calculations of the 
absorption line profile is only done for the line of sight from $z=0$ AU to 40 
AU in model i and iv and 37.5 AU in model ii and iii.

Potentially the adiabatic models may show absorptions with RV 
$> 0\,{\rm km}\,{\rm s}^{-1}$ and some synthetic spectra show weak traces of 
an absorption at the base of the red wing of the emission line. It must be 
noted that the chosen line of sight geometry for the synthetic absorption 
profiles avoids the jet backflow region. The backflow region would be clearly
visible for inclined line of sights or more extended emission regions.

Further we disregard in the line profile calculation the ionization problem 
discussed in section \ref{sec_ionization} assuming that all Ca-atoms are 
singly ionized \element[+]{Ca}. This is questionable particularly for the 
adiabatic jet models.

\begin{figure}
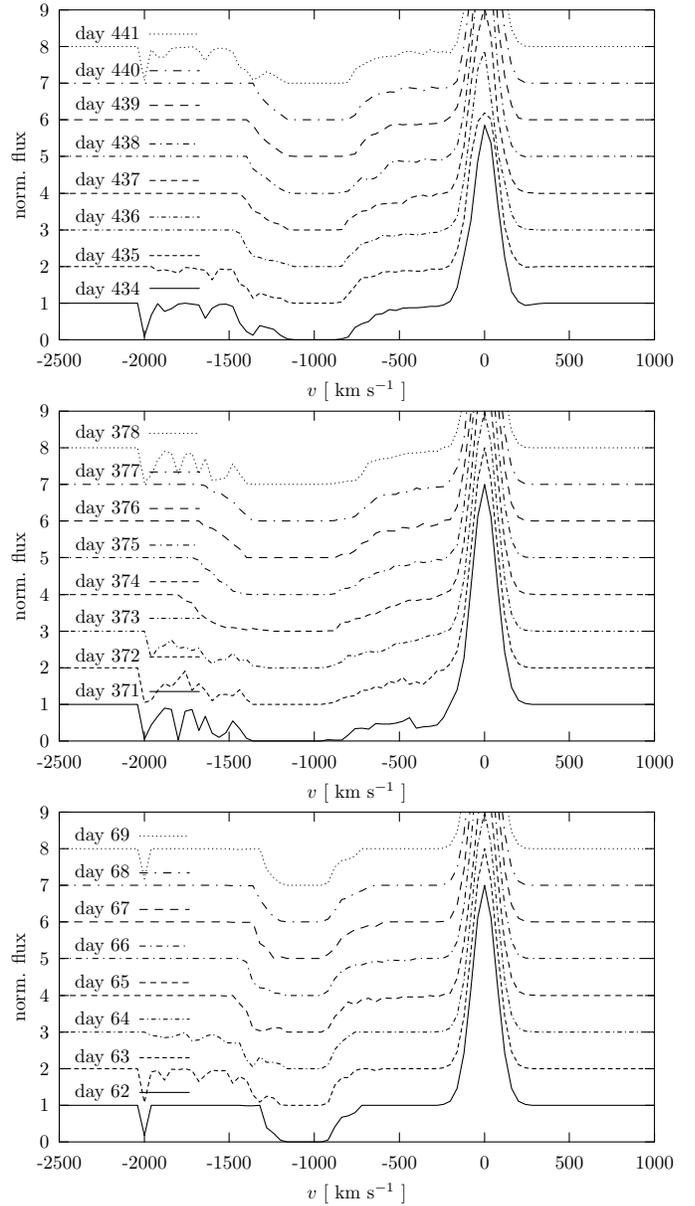

   \resizebox{\hsize}{!}{\includegraphics{line_time_NV2.0.eps}}

   \resizebox{\hsize}{!}{\includegraphics{line_time_NV2.3.eps}}

   \resizebox{\hsize}{!}{\includegraphics{line_time_NV3.0.eps}}
   \caption{Sequence of absorption line profiles for eight consecutive days of 
     model i (top), model iv (middle) and model i with cooling (bottom)}
   \label{line_time_NV2.0}
\end{figure}

With these restrictions we obtain in the simulations absorption line profiles 
which are shown in Fig. \ref{line_time_NV2.0}. There, 
sequences of the synthetical absorption line profiles are plotted for eight 
consecutive days for the adiabatic models i--iv. 

There exists quite some agreement between the synthetic profiles and the 
observations. The simulations produce the detached, broad absorption component 
as observed for times where no new high velocity component was ejected during 
the previous days in the jet nozzle. 

Not well reproduced is the strength of the high velocity absorption components 
due to the jet pulses. They are too weak by a substantial amount. At least one 
can see that the equivalent width and the depth of the high velocity 
components increases as the density in the jet pulse increases from 
$n_{\rm pulse}= 1.25, 2.5, 5$ to $10\times 10^6\,{\rm cm}^{-3}$ for models i, 
ii, iii and iv respectively. Further it is visible, that the high velocity 
absorption components of the higher density pulses require a longer time scale 
to be decelerated than the low density pulses. For example in Fig. 
\ref{line_time_NV2.0} (top), these components can be detected in only two 
consecutive days while they are present during three days in Fig. 
\ref{line_time_NV2.0} (middle).

Also a shallow component between the absorption trough and the emission line 
is visible in all models. The absorption is stronger for the adiabatic jets 
with higher pulse densities. The strength of this absorption is anticorrelated 
with respect to the high velocity component as in the observations. 

\subsection{Variations of the jet absorptions for the model with cooling}

Also for calculations of the jet absorptions from model i with cooling we have 
excluded the line of sight section through the ambient medium in front of the 
jet. This avoids the narrow, saturated absorption component in the emission 
component. Thus the calculation of the absorption profile was only performed 
from $z=0$ to 21 AU. 

The absorption for the jet model i with cooling behaves qualitatively similar 
to the adiabatic model. There is a broad jet absorption trough centered at 
about $\approx$ -1100 km s$^{-1}$, which is completely detached from the 
emission component. And again the high velocity absorptions are far too weak 
when compared with the observations. Comparing the models i with and without 
cooling seems to indicate, however, that they are a bit deeper and persist for 
a somewhat longer time in the model with cooling.

We like to note that it was also assumed for model i with cooling that all Ca 
is singly ionized. According the the model calculation the jet gas would be 
too high for \element[+]{Ca}, but the discrepancy between required temperature 
for \element[+]{Ca} and calculated temperature is not very large, in 
particular much smaller than for the adiabatic case. 

\section{Summary and discussion}

This paper describes pulsed jet models for parameters representative for
symbiotic systems. Two main types of simulations were performed, a small model 
grid for adiabatic jets with different jet pulse parameters and one simulation 
including radiative cooling. 

For the adiabatic jet models some quantitative differences are present for the 
different pulse parameters, but qualitatively the model results are very 
similar. 

Huge differences can, however, be seen between the adiabatic models and the 
model which includes radiative cooling. Compared to the adiabatic models the 
most important effects induced by radiative cooling are:  

\begin{itemize}
\item The energy loss through radiation lowers the pressure and the temperature
which leads to a much smaller radial extension of the jet, so that the cross 
section and the resistance exerted by the external medium is strongly reduced
leading to a significantly higher bow shock velocity.

\item The internal structure of the pulsed jet with cooling shows a well 
defined periodic shock structure with high density, low temperature knots 
separated by hot, low density beam sections. Contrary to this in the adiabatic 
jet the hottest regions have the highest densities while the density is low in 
cool regions. The temperature and density contrast is much more pronounced in 
the models with cooling.
\end{itemize}

These differences are generic for model results from jet simulations without
and with cooling. For the particular model i described here we can now quantify
these differences. 

\begin{itemize}
\item The jet radii are 10 AU for the adiabatic jet and 2.5 for the jet with 
cooling. These values are determined for model jets with the same axial 
length of 24~AU.

\item The bow shock velocity after 74 days is about 
$200\,{\rm km}\,{\rm s}^{-1}$ in the adiabatic model compared to 
$730\,{\rm km}\,{\rm s}^{-1}$ in the cooled jet.

\item The gas temperature in the adiabatic jet is everywhere (except for the
initial jet gas in the nozzle) higher than $T\approx 2\cdot 10^5$~K and goes 
up to $T\approx 10^7$~K in the high density shock regions. In the jet model 
with cooling the temperature is really low $T \approx 10^4$~K in the cool, 
high density knots. The hot beam regions are $T \gg 2\cdot 10^5$~K. 

\item The density contrast between cool and hot regions is about a factor of
30 in the adiabatic jet, but a factor of 1000 in the jet with radiative
cooling. Note that high density regions are hot in adiabatic models, while
they are cool in the models with cooling.  
\end{itemize}

These huge differences show clearly that adiabatic jet models are far from 
reality if radiative cooling is indeed important. Therefore jet models with 
radiative cooling should be investigated in much more detail for the high jet 
gas densities encountered in symbiotic systems. 

Unfortunately, the time scales of cooling processes are several orders of 
magnitude shorter than those of hydrodynamical ones. This is a large numerical 
problem which is demanding high computational costs. But as mentioned above, 
these costs are necessary to get new insights into the physics of jets in 
symbiotic stars. 

Until now, only two-dimensional simulations have been performed, also due to
high computational costs and memory demands of 3D-simulations. In those 
simulations, the red giant would be taken into account correctly, not as a 
ring, and its influences on the jet, the gravitation and the stellar wind, 
would not be symmetric anymore. This would result in the backflow being 
blocked only in a small segment, therefore the adiabatic jet should be more 
turbulent from the beginning.

As \citet{KoM} and \citet{KrC} have shown in their simulations, the numerical 
resolution determines, if one sees the transition from the laminar to the 
turbulent phase or not. Also in our simulations, this transition can be seen, 
implying that the resolution here is not too small. An increased resolution 
should not change the main result, the shrinking of the cocoon and the 
necessity of performing hydrodynamical simulations with cooling to understand 
the structure and emission of jets in symbiotic stars.

\subsection{Comparison with symbiotic systems}

Our model calculations make various predictions for jets in symbiotic systems. 
We focus here on the basic question whether the observations of symbiotic 
systems support more the adiabatic jet model or the jet model with cooling. 

For MWC 560 the spectroscopic observations show strong absorptions from 
low ionization species such as \ion{Ca}{ii}, \ion{Fe}{ii} and \ion{Na}{i}.
Absorptions from neutral or singly ionized metals are only expected for cool 
gas with temperatures around $T\approx 10000$~K or lower. The strength of 
these absorptions suggest that a substantial fraction of the jet gas must be 
in this cool state as in the jet model with cooling. This is incompatible with 
the hot gas temperatures $T \gg 10^5$ present in adiabatic models. The 
presented adiabatic jet models are not able to produce low temperature regions.

The propagation of the jet bow shock was well observed for the outburst in CH 
Cyg. The jet expansion was linear and fast 
$\approx 1500\,{\rm km}\,{\rm s}^{-1}$ producing after one year a narrow jet 
structure with an extention well beyond 300~AU. The estimated jet gas 
temperature was found to be below 10000~K. This is again in very good 
agreement with the jet model with cooling. According to our models an 
adiabatic model is expected to produce a jet with a lower expansion velocity, 
a broader jet beam emission and most importantly only high temperature jet 
gas. For CH Cyg it must be said that the initial jet conditions ``at the 
nozzle'' are not known. 

Less clear is the situation in R Aqr. Although jet features can be traced down 
to a distance of less than 20~AU, it is not clear whether the jet gas is cool 
or hot. A proper motion for the emission features in the range 36 to 
\mbox{240 km s$^{-1}$} has been derived. This would be more in the range of 
the gas motion of the adiabatic models. However it is not clear from where the 
observed emission originates, from the jet gas or the surrounding gas excited 
by shocks induced by the jet propagation. The small jet opening angle of 
15$^{\circ}$ favors more a jet with cooling, but a strong disk wind or a steep 
density gradient for the circumstellar gas may also help to enhance the 
collimation of an adiabatic jet.

Thus, the comparison between model simulation and observations suggest 
strongly that radiative cooling is important for jets in symbiotic systems. 
The fact, that the model with cooling describes the observations better than 
those without cooling, means that the cooling time scale in the jet of 
symbiotic systems must be shorter than the propagation time scale (i.e.
expansion time scale of the adiabatic gas) which is of the order of hundred 
days. This condition sets a lower limit for the density inside the jet of 
about $n \approx 10^6~{\rm cm}^3$ ( $T \approx 10^4$ K ) which is consistent 
with our initial parameters.

\subsection{The synthetical absorption line profiles}

Synthetic line absorption profiles are presented in this work, based
on hydrodynamical calculations of pulsed jets. These ``theoretical''
profiles are compared to observations of the jet in the symbiotic
system MWC 560. 

An important point is that the gas temperature in the adiabatic jet is 
everywhere far too high to produce the low ionization absorptions seen in the 
observations. 
Even in the model simulations including radiative cooling the 
gas temperature in the jet is too high  for producing the
observed \ion{Ca}{ii} line strengths. Thus, more efficient cooling 
of the jet gas is required. This can be achieved with higher gas densities
in the model jet outflow or perhaps with higher resolution 
of the hydrodynamical calculation, which may then be
able to resolve high density small scale clumps. A third 
solution could be the adding of further cooling processes which would be 
important at temperatures below $10^4$ K. Additional modeling is
required to investigate which of these solutions is more likely. 

Disregarding the ionization problem, we calculated the synthetic jet 
absorption line structure. This line structure represents essentially 
a projection of the gas radial velocity along the line of sight, which 
is more or less parallel to the jet. 

A success of our computations is that the basic structure of the jet
absorption in MWC 560, which is the broad, detached component, can be
well reproduced. The mean velocity and the velocity width is in good
agreement with the observations. Surprisingly, both adiabatic jet
models and the model with cooling provide qualitatively 
similar absorption profiles for the jet beam and the transient
high velocity components. Here are the merits of the adiabatic models which 
enable us to perform parameter studies which would be highly expensive for 
models with radiative cooling. However, as we disregard the ionization 
equilibrium
we should more correctly speak of the projected RV distribution of the gas.  

Also the temporal evolution of the highest velocity components of the transient
jet pulses follows the observed behavior. Not well reproduced by our
simulations are the strengths of the high velocity components. They
are far too week in our simulations. This suggests that higher gas
densities are required for the jet pulses. 

From the presented comparison between the synthetic and observed jet
absorption line structure we conclude that the general direction of
our modeling is correct. Due to the high temperature
the adiabatic simulations are inadequate to explain the observed low
ionization absorptions observed. However, disregarding the
temperature, they provide an easy to calculate mean to explore the
projected velocity distribution of the gas in a pulsed jet along the
line of sight. Our current jet simulation with radiative cooling
solves the ionization problem not fully, as the gas temperature is
still somewhat too high. However, the discrepancy is rather small and
may be solved with calculations having higher gas
densities in the jet in order to enhance the efficiency of gas
cooling. Higher gas densities are also required for the strengths of
the absorption in the transient high velocity components. 

With our comparison between synthetic and observed line profiles
we have demonstrated the huge diagnostic potential of the 
spectroscopic observations of the jet absorption profiles in MWC 560. 
The information, which can be extracted from these observations, 
is unique for astrophysical jets and is therefore a
most important source for a better understanding of astrophysical jets, in
particular for the detailed investigation of the propagation and
evolution of small scale structures originating in the jet acceleration 
region.  

The shape of the high velocity components is in all simulations discrete. This
could be a result of the assumed rectangular velocity and density steps for the
pulses. A Gaussian form possibly could smooth the components in the absorption 
line profiles.

A point where these simulations can be further improved is the elapsed time 
and therefore the calculated length of the resulting jet. We only simulated 
the jet until it reached a length of 50 AU. At this point, however, the 
density of the environment has decreased to the density in the jet nozzle. A 
transition of the initially underdense jet ($n_{\rm jet}/n_{\rm wind} <$ 1) 
towards an overdense jet will occur, which should result in changed 
kinematics. A direct effect on the absorption line profiles should be a 
decreased density of the shocked ambient medium, which was artificially cut 
out in the calculation of the profiles in this paper. After extending the 
computational domain, this manipulation should not be necessary. Simulations 
with larger domains are currently being carried out.

\begin{acknowledgements}
Parts of this work were supported by the Deutsche Forschungsgemeinschaft (DFG).
One author (M.S.) wants to thank J. Gracia and M. Krause for many fruitful 
discussions and the High Performance Computing Center Stuttgart for 
allowing to perform the expensive computations. We acknowledge the improving 
comments and suggestions by the referee, Garrelt Mellema. 
\end{acknowledgements}

\end{document}